\documentclass[3p]{elsarticle}

\makeatletter
\def\ps@pprintTitle{
	\let\@oddhead\@empty
	\let\@evenhead\@empty
	\def\@oddfoot{\centerline{\footnotesize{Accepted for publication in Results in Control and Optimization. Published version can be accessed at: \url{https://doi.org/10.1016/j.rico.2021.100078}}}}
	\let\@evenfoot\@oddfoot}
\makeatother

\usepackage{lineno,hyperref}
\usepackage{amsmath}
\usepackage{lipsum}
\usepackage{caption}
\usepackage{subcaption}

\begin{document}

\begin{frontmatter}
\title{Mitigating Biological Epidemic on Heterogeneous Social Networks}

\author{Jagtap Kalyani Devendra}
\ead{jagtap19@iiserb.ac.in}
\author{Kundan Kandhway}
\ead{kundankandhway@iiserb.ac.in}
\address{Department of Electrical Engineering and Computer Science,\\
Indian Institute of Science Education and Research Bhopal, Madhya Pradesh, 462066, India}
\date{}
\begin{abstract}
Recent Covid-19 pandemic has demonstrated the need of efficient epidemic outbreak management. We study the optimal control problem of minimizing the fraction of infected population by applying vaccination and treatment control strategies, while at the same time minimizing the cost of implementing them. We model the epidemic using the degree based \textit{Susceptible-Infected-Recovered} (SIR) compartmental model. We study the impact of varying network topologies on the optimal epidemic management strategies and present results for the \textit{Erd\H{o}s-R\'{e}nyi}, \textit{scale free}, and real world networks. For efficient computational modeling we form groups of groups of degree classes, and apply separate vaccination and treatment control signals to each group. This allows us to identify the degree classes that play a significant role in mitigating the epidemic for a given network topology. We compare the optimal control strategy with non optimal strategies (constant control and no control) and study the effect of various model parameters on the system. We identify which strategy (vaccination/treatment) plays a significant role in controlling the epidemic on different network topologies. We also study the effect of the cost of vaccination and treatment controls on the resource allocation. We find that the optimal strategy achieves significant improvements over the non optimal heuristics for all networks studied in this paper. Our results may be of interest to governments and healthcare authorities for devising effective vaccination and treatment campaigns during an epidemic outbreak.
\end{abstract}


\end{frontmatter}

\section{Introduction}

As of August 2021, SARS-CoV-2 has infected more than $200$ million people and more than $4$ million people have died because of the infection throughout the world \cite{14}. Hence, developing effective mitigation strategies through vaccination, treatment, lockdown and quarantine etc. have become crucial for managing this global pandemic; and has gained significant importance recently. The study of epidemic processes has been an area of interest for many years in various fields (e.g. computer science, epidemiology, and medicine). In recent years, various mathematical models of epidemic propagation and control have been designed for understanding how certain epidemic propagates in a given population and how to control the spread of such epidemic by devising various optimal control strategies. We can explain various epidemic processes such as the spread of infectious disease, information diffusion in a given population, spread of rumors, spread of malware/virus in the internet etc. using these mathematical models. These epidemiological models are of two types: deterministic and stochastic \cite{13}. In the deterministic models, the epidemic process is described as a system of differential equations; whereas, in the stochastic models, the epidemic process is described using the Markov process or stochastic differential equations.

In this paper we will focus on a deterministic model and we will formulate an optimal control problem for mitigating an arbitrary epidemic on a heterogeneous social network. Social networks are formed by individuals (people) as `nodes' and interaction between two individuals as `links'. We use the degree based compartmental Susceptible-Infected-Recovered (SIR) epidemic model. The three compartments of the SIR model are as follows: `Susceptible' are those individuals who have not contracted the disease yet but can potentially get infected if they come in contact with infected individuals. `Infected' individuals are those who have the disease and can pass on the infection to susceptible individuals. `Recovered' individuals are those who have recovered from the infection either on their own or due to undergoing treatment, or have died because of the infection, or have gained immunity after taking the vaccine. Suppose an epidemic breaks out in certain part of the network and the pathogen for the disease is known, then vaccination can effectively contain the epidemic by immunizing the population. Furthermore, treatment hastens the recovery of infected individuals. In this paper, we will employ optimal vaccination and treatment control strategies for containing the epidemic.

\subsection{Related Works} 

In this section we will discuss some of the related works in the literature. We broadly divide these works into four different classes: review papers, epidemic diffusion without application of control/optimization, epidemic control on homogeneously mixed population, and epidemic control on heterogeneous networks.

Readers who are interested in learning the basics of various epidemic models for homogeneously mixed population may refer to the reviews presented in \cite{1,26}. Fully mixed population models assume that every individual in the population has equal probability of contacting every other individual. However, in the real world only a small fraction of the total population is in contact with each other. In recent years, many researchers have studied the epidemic processes on complex networks. The work in \cite{12} provides a comprehensive review of the research on epidemic processes on complex networks discussing the theoretical approaches as well as their limits and assumptions. Authors in \cite{13} have reviewed and analyzed some of the most popular deterministic and stochastic epidemiological models for both homogeneously and heterogeneously mixed population. Results on optimization and control of epidemic dynamics were also discussed and outline for further exploration of the field was presented. The work in \cite{phase_transition} provides a concise review of epidemic models that exhibit phase transition from disease free to active stationary phase (where a fraction of population is infected) in heterogeneous networks.

In the following we discuss literature on epidemic diffusion on complex networks where optimization or control were not applied. Authors in \cite{6} have studied various epidemiological models and the respective epidemic thresholds on heterogeneous networks. They found that network topology plays a significant role in the overall behavior of epidemic spreading, for example, large connectivity fluctuations strengthens the incidence of epidemic outbreak. In \cite{7}, the author provides the exact solution for SIR model on networks of various kinds using percolation and generating function methods. A solution for the spread of STDs on bipartite graph of men and women is also presented. The work in \cite{mobility_sir} presents a mobility based SIR model for complex network that is used to forecast ongoing Covid-19 pandemic cases at country and regional level.

The work in \cite{8} analyzes the behavior of SIR and SIS epidemic models on complex networks with different levels of degree correlations. Authors present many useful results, for example, when vertices of the network are randomly drawn from a specified degree distribution $ p(k) $, the epidemic threshold vanishes for scale free networks with characteristic exponent $ 2 < \alpha \leq 3 $ in the limit of large network size. Authors in \cite{23} have studied the epidemic dynamics on finite size scale free network with soft and hard connectivity cutoffs. They have shown that the highly heterogeneous nature of scale free networks does not allow the use of homogeneous approximation even for relatively small networks. In all of these papers authors have studied the dynamics of epidemics on complex networks without the intervention of any external control signal. In contrast, we provide optimal strategies to control epidemic on various heterogeneous networks (\textit{Erd\H{o}s-R\'{e}nyi}, \textit{scale free} and a real world network).

There is a huge literature on epidemic control for homogeneously mixed population. In \cite{2}, authors have studied optimal control of Kermack-McKendrick epidemic model with linear cost function, and vaccination as external control. Dynamic Programming and Pontryagin's Maximum Principle were used to solve the problem. Similarly, the work in \cite{3} computed optimal vaccination schedules for mitigating the epidemic using dynamic programming techniques. In \cite{25, 27}, authors have considered standard epidemic models such as SIR, SIRS, SEIR etc. and have used Pontryagin’s maximum principle to characterize optimal levels of vaccination and treatment control signals. More recent works \cite{non_parma_intervention, social_dist} have also employed non-pharmaceutical interventions (e.g. social distancing, home quarantining etc.) for optimal control of an epidemic.
 
In \cite{20}, authors have investigated dynamics of seasonal flu-like epidemics with and without control strategies. They studied the impact of optimal control strategies as a function of various model parameters using numerical simulations. In \cite{5} authors have devised pre-outbreak and during outbreak vaccination strategies for SARS epidemic. Latent (traced and untraced), isolated and dead individuals were also modeled. The study concluded that if vaccines are available, vaccination programs should be implemented as soon as the SARS outbreak is detected. Isolation should be made effective and identification of latent individuals through contact tracing was found to be crucial for mitigating SARS epidemic.

The usual mitigation strategy is to control susceptible and infected individuals by employing vaccination, quarantine or treatment controls. The work in \cite{17} has used multiple such controls on modified Kermack-McKendrick epidemic model. The study concluded that multiple controls are more effective than single control in reducing the number of infected and susceptible population. The work in \cite{15} has formulated an optimal control problem for information dissemination on homogeneously mixed population using Maki-Thompson rumor model with fixed budget constraint. In the above literature authors have assumed a homogeneously mixed population and have employed one or two control signals. In contrast, in this paper we have used heterogeneous networks which models the real world interaction more closely. In addition, the increased size of dynamical system leads to computational challenges. We have used separate vaccination and treatment control signals for groups of high, medium and low degree nodes as discussed in Section \ref{sec:gr_of_gr_and_control_sig}.

We now discuss literature on biological and information epidemics on heterogeneous networks. In \cite{18}, authors have developed vaccination strategies for the disease (such as influenza) which spreads from city to city due to airplane travel using subpopulation SIR model. In \cite{16}, authors explore the effect of isolating high degree nodes (high degree nodes have many connections and hence play a significant role in the spread of epidemic) using complete and incomplete isolation. The authors have used a mean-field SIR model. They found that complete isolation is relatively harsh and has large economical cost. Whereas, incomplete isolation is mild and allows low degree nodes to rewire and save some economical cost. The work in \cite{24} proposes effective \textit{acquaintance immunization} strategy (i.e. immunization of a random acquaintance of a random node using vaccination) on the scale free network. Both these works do not involve any optimization and propose sub-optimal strategies. However, we formulate an optimal control problem and obtain a numerical solution. 

In \cite{22}, authors have formulated an optimal control problem for the SIR model and have computed solution for a five node network. The techniques employed in our work allows us to present the results for a much larger network. Authors in \cite{4} have studied heuristic lockdown policies (partial/complete) on two-age structured SIR model. The aim was to reduce the socio-economic costs while taking into account constraints such as limited hospital capacity. The work in \cite{nonmassive_immu} constructs an optimal non-massive immunization strategy to contain SIS epidemic on complex networks, however, without employing the theory of optimal control.

We next discuss some literature on control of non-biological epidemics. The work in \cite{9, 10} discuss information diffusion on social networks and devise strategies for running optimal campaigns. Campaigns could be political, for product marketing, for spreading awareness about a social cause etc. In \cite{9}, authors have used Susceptible-Infective (SI) model for information diffusion on social network with fixed budget constraint using `direct' and `word-of-mouth' control signals. In \cite{10}, authors have formulated an optimal control problem for information dissemination on a heterogeneous network using `direct recruitment' of susceptible nodes to infected class. Optimal control problem is solved numerically using forward-backward sweep algorithm. The work in \cite{21} devises incentive strategies for viral marketing of a product. Authors have used `direct incentives' and `referral rewards' as control strategies. The optimal strategy takes a simple structure in which the seller needs to deploy these incentives in at the most two distinct time period. The simplistic structure makes these strategies easier to implement in practice.

Our contribution with respect to above literature is as follows: In \cite{9,10,21} authors have analyzed information epidemic. The nature of optimal control problem changes completely in biological epidemic. In information epidemic we try to maximize the spread whereas, in biological epidemic we try to minimize the spread. This also changes the control signals that are employed in the two class of problems. In addition, works such as \cite{10}, applies one control signal to each degree class. This is computationally expensive and impractical to implement in practice. Suppose a network comprises of $ 100 $ degree classes, then such a solution requires implementing $ 100 $ different optimal policies for each degree class. In real world network, knowing actual degree of every individual is not possible and since there are too many controls, precise implementation of control policies for every degree class is very difficult. We have circumvented above problem by grouping of degree classes and thereby effectively reducing computational complexity which will be discussed in Section 2.2.

The key contributions of this paper are as follows:
\begin{itemize}
\item We use \textit{vaccination} and \textit{treatment} control signals for mitigating the biological epidemic on degree based SIR compartmental model. We have modified the standard epidemic model by grouping the degree classes. This reduces the computational complexity of the model while still capturing the evolution of states (susceptible, infected, and recovered) with negligible error.

\item We compare the effect of optimal control strategy with heuristics such as constant control and no control strategies to quantify the effectiveness of optimal strategies against such heuristics. We study the spread of epidemic on different network topologies (\textit{scale free},\textit{ Erd\H{o}s-R\'{e}nyi}, and real world networks) and identify which control signal plays important role in mitigating the epidemic as the degree distribution of the network changes. We explore the effect of various model parameters on the resource allocation. 

\item The grouped degree classes of the SIR compartmental model are further amassed into three groups: \textit{Low, Medium} and \textit{High} degree groups. We apply separate vaccination and treatment control signals on each of these three groups. This allows us to identify important groups to be targeted for effective epidemic management. The results show that the importance of low, medium and high groups changes as degree distribution of the network changes---high degree group assumes maximum importance in the scale free network whereas medium degree group assumes maximum importance in the Erd\H{o}s-R\'{e}nyi network. It is not possible to draw such insights for biological epidemic management from formulations already existing in the literature.
\end{itemize}

This paper is divided into following sections: Section 2 describes the model and controls used in this study as well as the formulation of optimal control problem. Section 3 contains the discussion of the experimental set-up i.e. numerical techniques used to solve the optimal control problem, networks used in this study, default model parameters and heuristic strategies. In Section 4 we discuss resource allocation among groups and control strategies and identify important groups and control strategies for containing the epidemic. We also study the effect of varying various model parameters. Results for both synthetic and real world networks are presented in this section. Finally, Section 5 concludes the paper. 

\section{System Model and Problem Formulation}
In this section we first briefly discuss the uncontrolled system---the standard degree based compartmental SIR epidemic model. We then adapt the standard model by grouping the degree classes, and apply control signals to the groups of groups of degree classes. Finally we formulate the optimal control problem.

We have used the SIR model for studying the spread of disease on heterogeneous social network \cite{8}. We have used static social networks in this work. The degree of a node (individual) is the number of neighbors he/she has. Unlike in homogeneously mixed population where every individual in the population has equal chances of interacting with every other individual; in heterogeneous network, contacts (neighbors who are connected by a link) are fixed and is a small subset of the whole population. The nodes in the network follow arbitrary degree distribution, $ {p}_{k} $. All nodes with degree $ k $ are said to be in degree class $ k $. The set of all possible degree classes in the network is denoted by $ \mathcal{K} = \lbrace K_{min},..,K_{max}\rbrace $ where, $ K_{min} $ and $ K_{max} $ are the minimum and maximum degrees in the network. There are a total of  $ |\mathcal{K}| = K_{max} - K_{min} +1 $  degree classes in the network \cite{8}.

\subsection{Uncontrolled System}
Nodes (individuals) in the network could be in one of the three possible states: susceptible, infected, and recovered (as explained in Section 1). We assume that the epidemic breaks out in the giant component of the network. The epidemic starts at $ t = 0 $ and ends at $ t = T $ (i.e. $ T $ is the epidemic duration). Let $ s_{k}(t), i_{k}(t) $ and $ r_{k}(t) $ be the fraction of susceptible, infected and recovered nodes in degree class $ k $ at time $ t $. Infected nodes come in contact with other nodes at an average rate of $ \beta $ (known as the spreading rate) and infected nodes recover (or die) at an average rate of $ \gamma $ (also known as the recovery rate) per unit time. The model assumes that there is a small fraction of infected nodes, $ i_{0} $ in every degree class $ k $, which kick-start the epidemic. That is, $ i_{k}(0) = i_{0}; \forall k \in \mathcal{K}. $

All nodes in the same degree class behave the same way \cite{19}. If the `susceptible' node \textit{A} changes its state to `infected' after coming in contact with its neighbor \textit{B}, then neighbor \textit{B} must have caught the infection from one of its neighbors other than \textit{A}. Therefore, the probability of node \textit{B} being infected is given by $ i_{k} $ but here, $ k $ is the ``excess degree" of node \textit{B} (which is one less than the actual degree). Excess degree distribution is given by $ {q}_{k} = (k+1){p}_{k+1}/<k> $, where, $ <k> = \sum_{k \in\mathcal{K}} k{p}_{k} $ is the mean degree of the network \cite{19}. The average probability that the neighbor is infected is thus given by $ \sum_{k \in \mathcal{K}} {q}_{k}i_{k}(t) $. Therefore, the rate of change of susceptible, infected and recovered nodes in degree class $ k $ is given by \cite{19}:
\begin{subequations}\label{eqn:1} 
     \begin{align}  
     \dot{s}_{k}(t) = -\beta ks_{k}(t)\sum_{l \in \mathcal{K}} {q}_{l}i_{l}(t) \label{subeqn:1a} \\
    \dot{i}_{k}(t) = \beta ks_{k}(t)\sum_{l \in \mathcal{K}} {q}_{l}i_{l}(t) - \gamma i_{k}(t) \label{subeqn:1b} \\
    \dot{r}_{k}(t) = \gamma i_{k}(t) \label{subeqn:1c}
     \end{align}
\end{subequations} 
It is worth noting that RHS of the above three equations add up to zero and hence dependent. Any two of the three equations are independent and enough for complete modeling of the SIR epidemic process. 

\subsection{Grouping of Degree Classes}
The standard model in Eq. (\ref{eqn:1}) requires two differential equations for each degree class in the network. This is computationally expensive as $ 2|\mathcal{K}| $ differential equations have to be solved each time for computing the epidemic evolution. We propose a simple method to reduce the computational complexity of the epidemic model in the following. It should be noted that epidemic evolution has to be computed multiple times to compute the optimal control. Hence reducing the computational complexity of the epidemic model is useful. We form $ Z $ groups of total $ |\mathcal{K}| $ degree classes. The motivation for grouping the degree classes is as follows. States of the nodes of similar degrees evolve in a similar manner, i.e., evolution of states (susceptible, infected and recovered) of nodes with degree $ k $ is not much different from that of nodes with degree $ k+1 $. We can take advantage of this fact for reducing the computational complexity of the epidemic model by grouping these similar degree nodes. With this trick we only need to solve $ 2Z $ differential equations to capture the epidemic evolution.

Fig. 1 shows the combined (for all three states: susceptible, infected and recovered) relative error achieved as the number of groups $ Z $ is varied for both scale free and Erd\H{o}s-R\'{e}nyi networks. The scale free and Erd\H{o}s-R\'{e}nyi networks originally have 100 and 45 degree classes respectively. From the graph we can see that for $ Z\approx 20 $, combined relative error is below $ 10^{-3} $ for both the networks. Therefore, $ 20 $ groups are enough to model the epidemic accurately, i.e., epidemic evolution of $ 20 $ groups is as good as the original degree based compartmental model. Thus, we are able to achieve $ 5 $ fold reduction in the computational effort for the case of scale free network. The set of all degree classes in group $z$ is denoted by $ \mathcal{\hat K}_{z}, 1 \leq z \leq Z $. We form $ Z $ groups such that $ \sum_{k \in \mathcal{\hat K}_{z}} {p}_{k} \approx 1/Z $. The grouped degree distribution $ \hat p_{z}$, the grouped excess degree distribution $ \hat q_{z} $ and the weighted mean degree $ \hat k_{z} $ for each $ z $ are given by:
\begin{subequations}\label{eqn:2}
	\begin{gather}
		\hat p_{z} = \sum_{k \in \mathcal{\hat K}_{z}}{p}_{k} \label{subeqn:2a} \\
		\hat q_{z} = \sum_{k \in \mathcal{\hat K}_{z}}{q}_{k} \label{subeqn:2b} \\
		\hat k_{z} = \sum_{k \in \mathcal{\hat K}_{z}}k{p}_{k} / \sum_{k \in \mathcal{\hat K}_{z}}{p}_{k} \label{subeqn:2c}
	\end{gather}  
\end{subequations}
If $\hat s_{z}(t), \hat i_{z}(t)$ and $\hat r_{z}(t)$ denote the fraction of susceptible, infected, and recovered nodes in group $z$ at time $t$, the grouped uncontrolled system is defined as follows:
\begin{subequations}\label{eqn:3}
	\begin{gather}
		\dot{\hat s}_{z}(t) = -\beta \hat k_{z} \hat s_{z}(t)\sum_{l = 1}^{Z} \hat q_{l}\hat i_{l}(t);    ~1 \leq z \leq Z. \label{subeqn:3a} \\
		\dot{\hat i}_{z}(t) = \beta \hat k_{z} \hat s_{z}(t)\sum_{l = 1}^{Z} \hat q_{l} \hat i_{l}(t) - \gamma \hat i_{z}(t);   ~1 \leq z \leq Z.\label{subeqn:3b} \\
		\dot{\hat r}_{z}(t) = \gamma \hat i_{z}(t);   ~1 \leq z \leq Z.\label{subeqn:3c}
	\end{gather}
\end{subequations}

\subsection{Groups of Groups and Control Signals}
\label{sec:gr_of_gr_and_control_sig}

\begin{figure}[t!]
\centering
  \includegraphics[width=.45\linewidth]{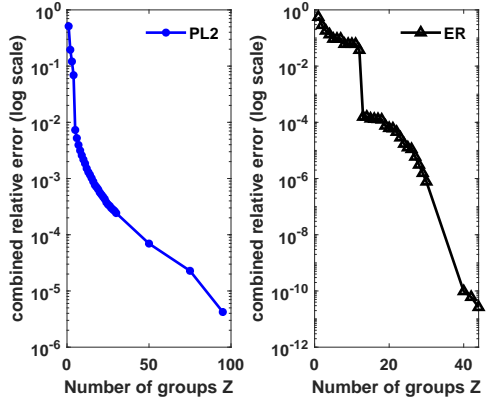}
  \caption{{\small Combined relative error vs. number of groups $ Z $ for PL2 and ER networks for the following parameters: Spreading rate $ \beta = 0.5$, recovery rate $\gamma = 0.25$, epidemic duration $T = 20$, initial fraction of seed nodes $i_{0} = 0.01. $}}
  \label{fig:Fig.1}
\end{figure}

To mitigate the effects of biological epidemic we use \textit{vaccination} and \textit{treatment} control signals as discussed below. We bundle $ Z $ groups further into $ M $ groups, denoted as $ \mathcal{G}_{m}, 1 \leq m \leq M $; and apply separate vaccination and treatment control signals to each of these groups $ \mathcal{G}_{m} $. We form $ M $ groups such that $ \sum_{z \in \mathcal{G}_{m}} \hat p_{z} \approx 1/M $. Since each group $ \mathcal{G}_{m} $ is controlled by one vaccination and one treatment control signal, $M$ different vaccination and treatment signals are used to control the system in Eq. (3a)-(3c). We denote $ M $ dimensional vaccination vector control function by $ \textbf{u} = (u_{1},..,u_{M}). $ The value $ u_{m}(t), 1 \leq m \leq M, $ denotes the rate of vaccination of susceptible individuals in group $ \mathcal{G}_{m} $ at time $ t $. Through the vaccination control signal we transfer individuals from the susceptible state to the recovered state directly. Since these individuals gain immunity, they neither catch the disease nor do they spread it further. The $ M $ dimensional treatment control signal is denoted by $ \textbf{v} = (v_{1},..,v_{M}). $  The value $ v_{m}(t), 1 \leq m \leq M, $ denotes the rate of treatment of infected individuals in group $ \mathcal{G}_{m} $ at time $ t $. Treatment control transfers individuals from the infected state to the recovered state at an increased rate compared to the natural recovery rate. Both control signals are continuous. Further, $ u_{m}(t) \geq 0 $ and $ v_{m}(t) \geq 0, 1 \leq m \leq M $.

\subsection{Controlled System}
Our aim is to compute optimal vaccination and treatment strategies while minimizing both the fraction of infected population, and the cost of vaccination and treatment. The objective function which would achieve this is:
\begin{gather}
J = \int_0^T \left\{ \sum_{z = 1}^{Z} \hat p_{z} \hat i_{z}(t) + b \sum_{m = 1}^{M} x_{m}u_{m}^{2}(t) +  c \sum_{m = 1}^{M} x_{m}v_{m}^{2}(t) \right\} dt \label{eqn:4}
\end{gather}
Note that $ T $ is the duration of the epidemic. Also,  $ x_{m} = \sum_{z \in \mathcal{G}_{m}} \hat p_{z} $ denotes the fraction of population in group $ \mathcal{G}_{m} $. The constants $b$ and $c$ are the weighing factors representing costs of vaccination and treatment respectively. The quantity $ \sum_{z = 1}^{Z} \hat p_{z} \hat i_{z}(t) $ measures instantaneous fraction of infected nodes at time $ t $. The cumulative infected population during entire duration of the epidemic outbreak is thus given by $ \int_0^T \sum_{z = 1}^{Z} \hat p_{z} \hat i_{z}(t) dt $. This represents the health state of the population during the epidemic outbreak. The quantities $ b \sum_{m = 1}^{M} x_{m}u_{m}^{2}(t) $ and $ c \sum_{m = 1}^{M} x_{m}v_{m}^{2}(t) $ capture the instantaneous resource consumption for vaccination and treatment respectively. The total resource consumption for vaccination and treatment during the entire duration of the epidemic is simply integration of the instantaneous resource consumption over the epidemic duration and is given by $ \int_0^{T} b \sum_{m = 1}^{M} x_{m}u_{m}^{2}(t)dt $ and $ \int_0^{T} c \sum_{m = 1}^{M} x_{m}v_{m}^{2}(t)dt $ respectively. The rate of change of susceptible and infected nodes in group $z$ after application of control signals is given by:
\begin{subequations}\label{eqn:5}
     \begin{align}
     \dot{\hat s}_{z}(t) = -\beta \hat k_{z} \hat s_{z}(t)\sum_{l = 1}^Z \hat q_{l} \hat i_{l}(t) - \hat s_{z}(t) u_{m}(t); ~z\in \mathcal G_m, ~1\leq m \leq M. \label{subeqn:5a} \\
    \dot{\hat i}_{z}(t) = \beta \hat k_{z} \hat s_{z}(t) \sum_{l = 1}^Z \hat q_{l} \hat i_{l}(t) - \gamma \hat i_{z}(t) - \hat i_{z}(t) v_{m}(t); ~z\in \mathcal G_m, ~1\leq m \leq M. \label{subeqn:5b} \\
     \hat s_{z}(t) + \hat i_{z}(t) + \hat r_{z}(t) = 1; ~1\leq z \leq Z \label{subeqn:5c} \\
     \hat i_{z}(0) = i_{0}, \hat s_{z}(0) = 1-i_{0}, \hat r_{z}(0) = 0; ~1\leq z \leq Z \label{subeqn:5d}
     \end{align}
\end{subequations}
The total fraction of susceptible, infected, and recovered nodes at any time $ t $ is given by:
\begin{subequations}\label{eqn:6}
     \begin{align}
     s(t) = \sum_{z = 1}^{Z} \hat p_{z} \hat s_{z}(t) \label{subeqn:6a} \\
     i(t) = \sum_{z = 1}^{Z} \hat p_{z} \hat i_{z}(t) \label{subeqn:6b} \\
     s(t) + i(t) + r(t) = 1 \label{subeqn:6c}
     \end{align}
\end{subequations}
As stated earlier, $ i_{0} $ is the initial fraction of infected nodes in each group $z$ which acts as a seed for spreading the disease. Since $ r(t) = 1 - s(t) - i(t)$, we need $ 2Z $ state equations to capture the spread of disease.

\section{Experimental Set-up}
In this section we will first discuss the numerical techniques used for solving the optimal control problem described in Eqs. (4-5). We will then discuss the synthetic networks used in this paper. Further, we will discuss the default model parameters, and finally the heuristic strategies used to compare with the optimal strategy.

\begin{figure}[t]
	\centering
	\includegraphics[width=.45\linewidth]{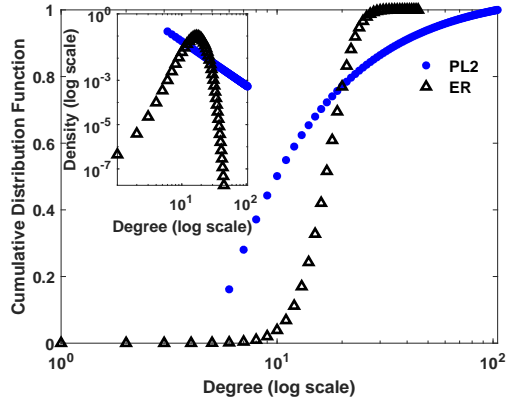}
	\caption{{\small Cumulative degree distribution and degree distribution (inset) of networks used in this study.}}
	\label{fig:Fig.2}
\end{figure}

\subsection{Numerical Techniques Used}
Nonlinear programming tools available in Matlab can easily be used to solve the constrained and unconstrained optimization problems. We have solved the discretized version of the optimal control problem, with cost function in Eq.(4), subject to the constraints in Eq. (\ref{eqn:5}) using Matlab's non-linear optimization solver \textit{fmincon()} as described in \cite{11}. The epidemic duration is sampled at $ \mathcal{N} $ equidistant points, $t_n, ~0 \leq n \leq \mathcal{N}-1$. The values of the control signals $ u_{m}(t_{n}) $ and $ v_{m}(t_{n}); ~0 \leq n \leq \mathcal{N}-1; ~1 \leq m \leq M $ are computed using the \textit{fmincon()} optimization routine. The computation for $ u_{m}(t_{n}) $ and $ v_{m}(t_{n}); \forall n, \forall m $ in the optimization routine starts with some initial guess and the function refines it until stopping criteria are met. We have used \textit{Heun's method} to discretize the system of ODEs in Eq. (\ref{eqn:5}). For the same number of sampling points, Heun's method is faster than Runge-Kutta method, but slower than Euler's method. However, it is less accurate than Runge-Kutta method, and more accurate than Euler's method. Taking into account the execution time, memory constraints, and accuracy, Heun's method is found to work well in practice.
 
\begin{figure}[t!]
\centering
  \includegraphics[width=.45\linewidth]{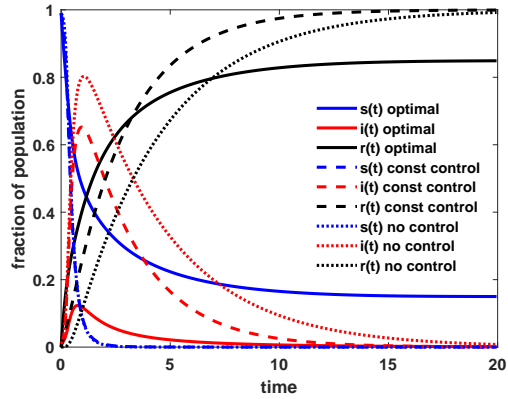}
  \caption{{\small Evolution of states for different strategies for PL2 network. Parameter values: Spreading rate $ \beta = 0.5$, recovery rate $\gamma = 0.25$, epidemic duration $T = 20$, initial fraction of seeds $i_{0} = 0.01. $}}
  \label{fig:Fig.3}
\end{figure}

\subsection{Synthetic Networks Used In This Study}
In this section we discuss the two synthetic networks used to present the results in this paper (the real world network is discussed later in Section 4.4). The first one is an Erd\H{o}s-R\'{e}nyi network, which is known to have Poisson degree distribution. The probability of finding a node with degree $ k $ is $ {p}_{k} = e^{-\lambda} \lambda^{k}/k!, ~\forall k \in \mathcal{K}$, where $ \lambda $ is the mean degree of the network. The second network used in this study is the scale free network which follows the power law degree distribution given by $ {p}_{k} = \nu k^{-\alpha}, ~ \forall k \in \mathcal{K} $, where $\nu$ is the normalization constant given by $ \nu = 1/\sum\limits_{k=K_{min}}^{K_{max}} k^{-\alpha} $. Here, $ \alpha $ is the power law exponent. For most real world networks $\alpha$ lies between 2 and 3 \cite{19}. We have chosen $ \alpha = 2 $ in this study. Henceforth, we will refer to the above two networks as ER and PL2 respectively.

The mean degrees of both networks are almost the same. They cannot be made equal because the minimum and maximum degrees are discrete quantities. The minimum and maximum degrees of both networks are as follows: $ K_{min}^{PL2} = 6, K_{max}^{PL2} = 105 $, with mean degree equal to $ <k>_{PL2} = 17.1818 $; and $ K_{min}^{ER} = 1, K_{max}^{ER} = 45 $, with mean degree equal to $ <k>_{ER} = 17.5 $. Thus, total number of degree classes in both the networks are, $ |\mathcal{K}^{ER}| = 45 $ and $ |\mathcal{K}^{PL2}| = 100 $ respectively. The probability degree distribution and the cumulative degree distribution for the above two networks are shown in the Fig. 2. We have formed $ Z $ groups of $ |\mathcal{K}| $ degree classes as discussed in Section 2.2. Fig. 1 shows combined relative error with respect to number of groups $ Z $. The value of $ Z $ for both networks is $ 21 $. We choose this value of $ Z $ because combined relative error goes below $ 10^{-3} $ for both the networks. Thus, we are able to achieve a five fold and a two fold increase in the computational efficiency by grouping of degree classes for PL2 and ER networks respectively. In the real world network it is often difficult to find the exact degree of the individual. Fig. 1 shows that even when the degrees of individuals are known at a coarser level, we can still use degree based compartmental models accurately.  

For the purpose of application of controls we have further amassed the $Z$ groups into $ M $ groups. For practical reasons we choose $ M = 3 $. Fewer number of control signals make the large scale practical implementation easy. We name these three groups as low $ \mathcal{G}_{1} $, medium $ \mathcal{G}_{2} $, and high $ \mathcal{G}_{3} $ degree groups. The resource allocated to these three groups in the optimal strategy signifies their relative importance in the epidemic management for a given network topology. Division of $Z$ groups into $M$ groups is done using the approach discussed in Section 2.3.

\subsection{Default Model Parameters}
The default model parameters are set to following values unless stated otherwise for the rest of this paper. The initial fraction of infected population, $ i_{0} $ is set to $ 0.01 $. The epidemic duration, $ T $ is set to $ 20 $. The spreading rate, $ \beta $ and the recovery rate, $ \gamma $ are set to $ 0.5 $ and $ 0.25 $ respectively. We choose these values of $ \beta $ and $ \gamma $ because uncontrolled system for these parameter values leads to large epidemic outbreak, which is the situation we would want to avoid. We have not set the upper bounds on the values taken by the control signals $ u_{m}(t) $ and $ v_{m}(t)$ because we are using quadratic cost function in Eq. (4). This will ensure that the vaccination and treatment control signals are bounded and take finite values. The instantaneous cost of applying vaccination and treatment control signals are $ bx_{m}u_{m}^{2}(t) $ and $ cx_{m}v_{m}^{2}(t) $ respectively, with constants $ b $ and $ c $ set to $ 0.25 $ and $ 0.5 $ respectively for all $m$. 

\subsection{Heuristic Strategies}
We have compared the effectiveness of the optimal strategy against two heuristics: \textit{constant (static) control} strategy and \textit{no control} strategy. We will demonstrate that the optimal control strategy works better than their non optimal counterparts in Section 4. For constant control strategy, we have set vaccination and treatment control signals to a constant value of $ \beta/2 $ and $ \gamma/2 $ respectively. Evolution of the fraction of susceptible, infected and recovered population for the three strategies are shown in Fig. 3 for PL2 network. The cumulative infected population for the optimal, constant, and no control strategies are $ 0.3810, 2.3277 $ and $ 3.9683 $ respectively.

\section{Results and Discussion} 
In this section we will discuss resource allocation among $ M $ groups and control strategies (vaccination and treatment) with the purpose of identifying important groups and control strategies for containing the epidemic on different network topologies. We will discuss the effect of varying the spreading rate $ \beta $ and the relative cost of control strategies on the objective function.

\subsection{Resource Allocation Among Groups and Strategies}
\begin{figure*}[t]
  \centering
  \subcaptionbox{Controls, ER}[.45\linewidth][c]{%
    \includegraphics[width=\linewidth]{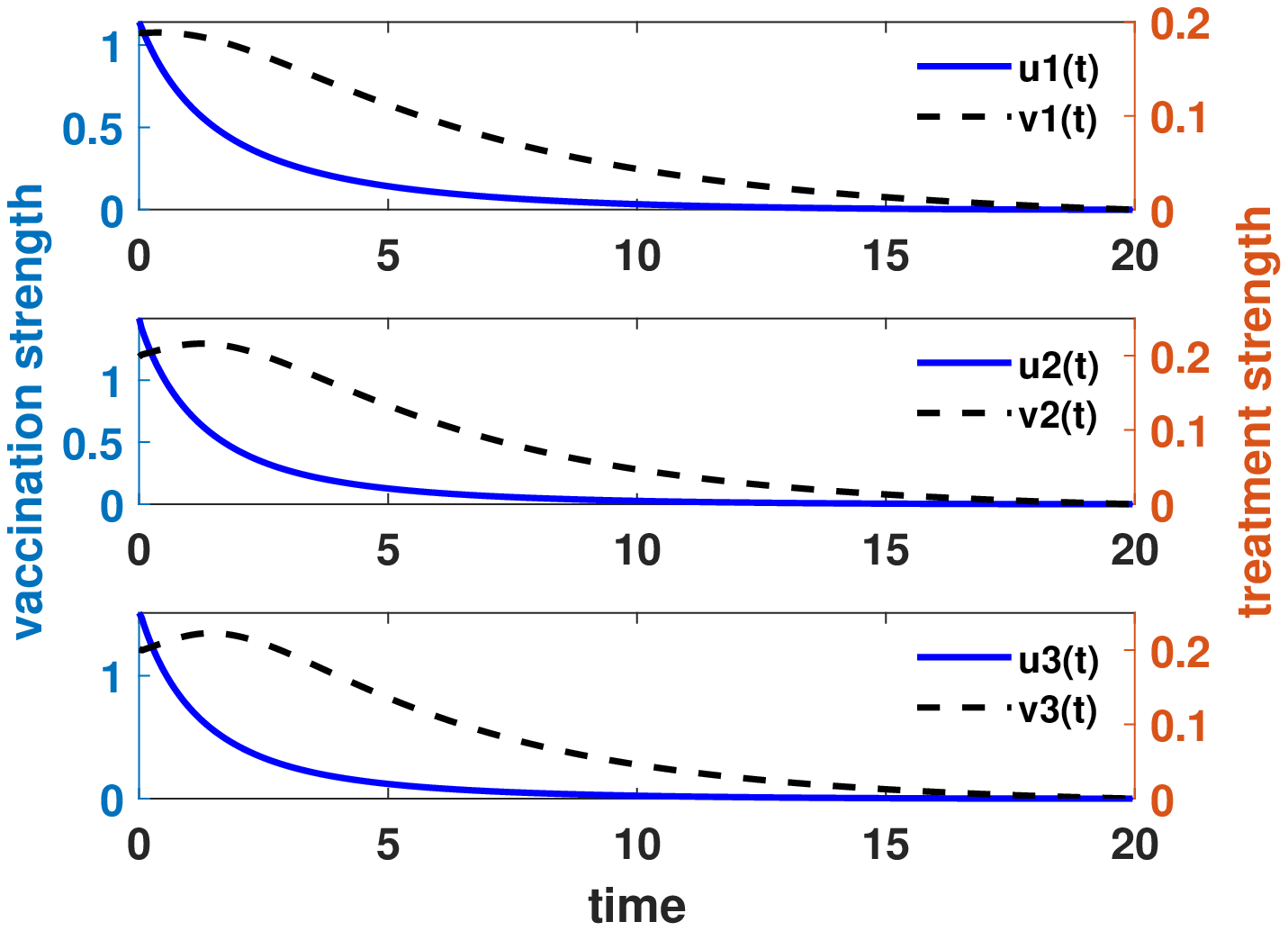}}
  \subcaptionbox{Controls, PL2}[.45\linewidth][c]{%
    \includegraphics[width=\linewidth]{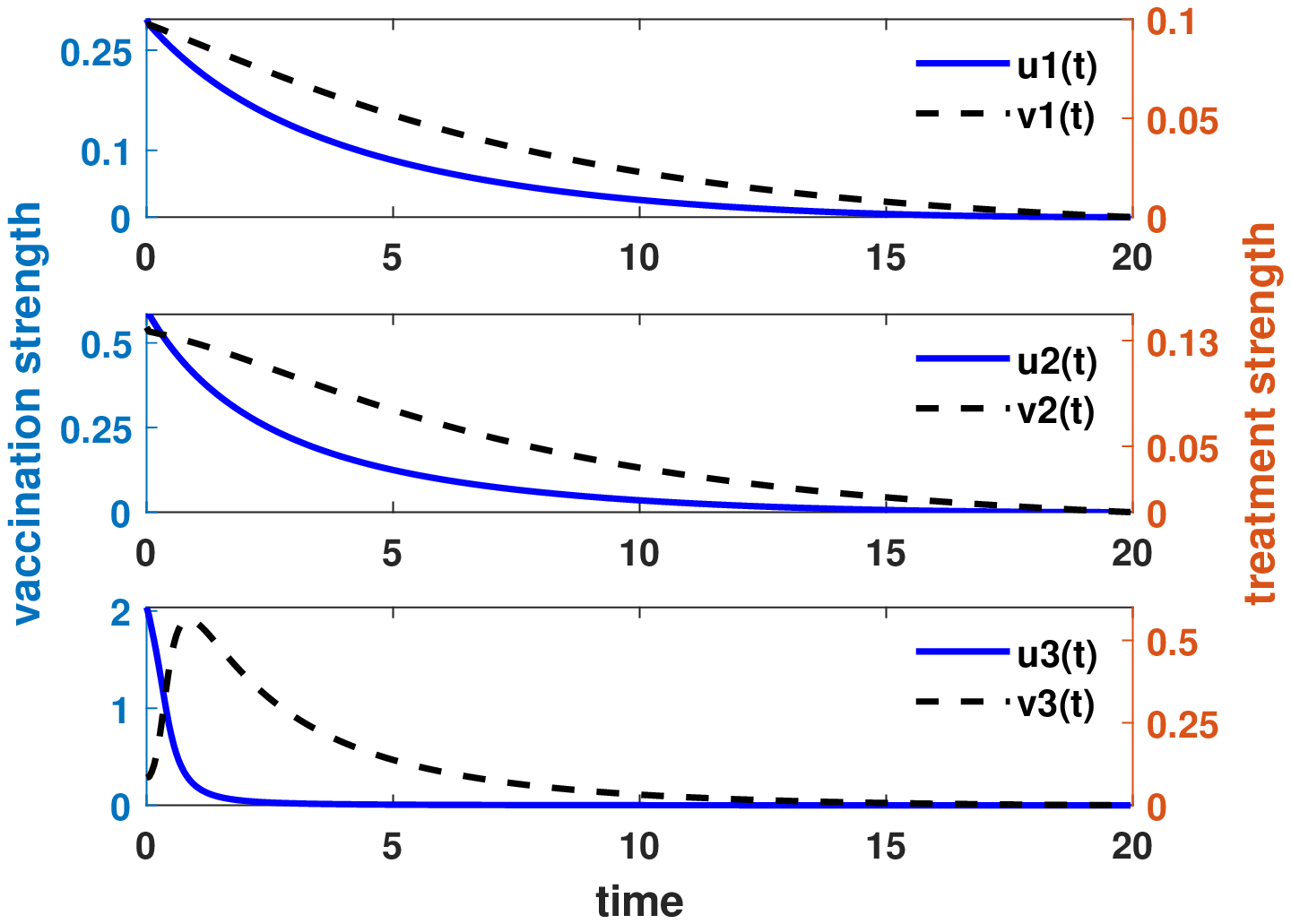}}
	\bigskip

  \subcaptionbox{$\% $ Resource Allocation , ER}[.45\linewidth][c]{%
    \includegraphics[width=\linewidth]{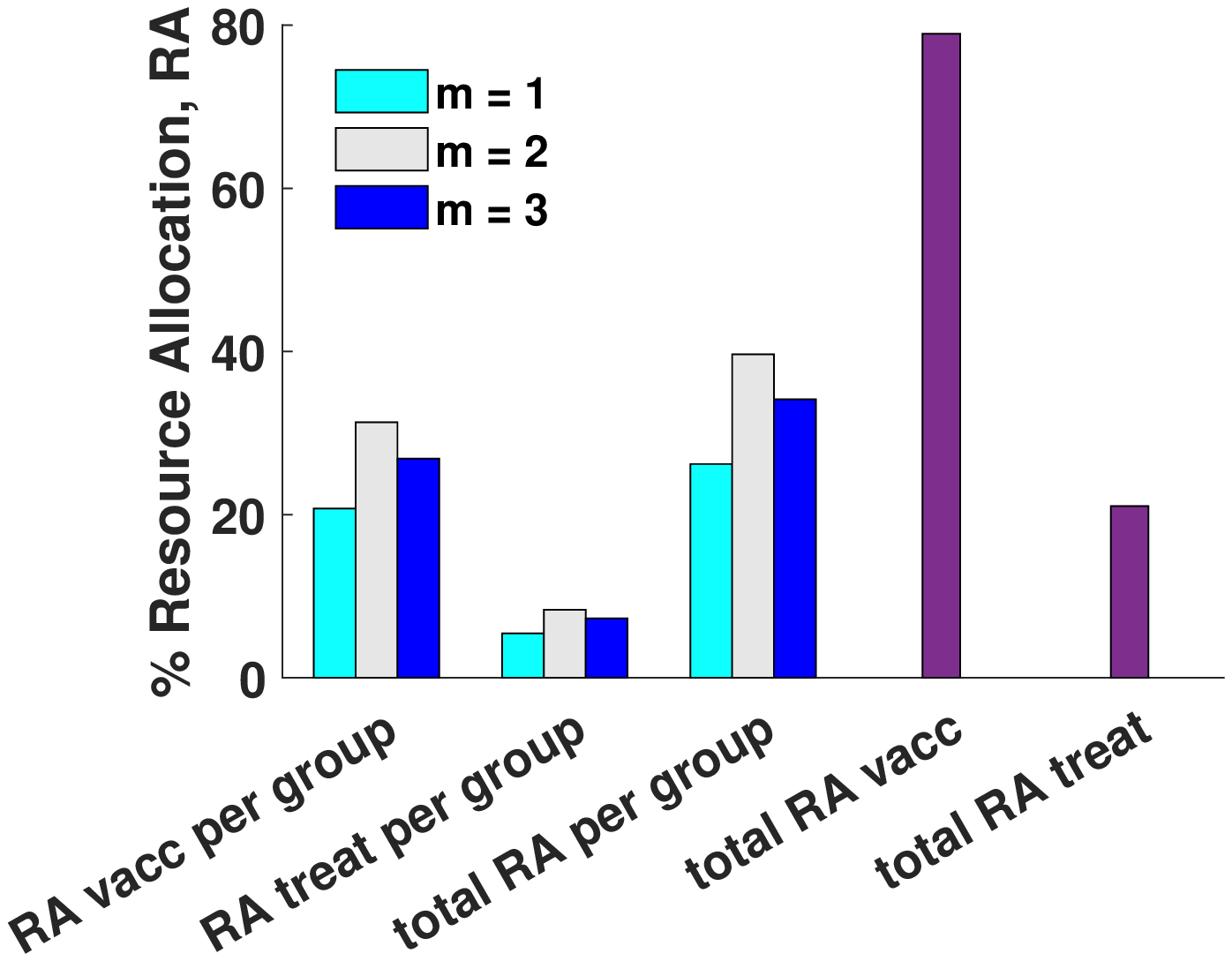}}
  \subcaptionbox{$\% $ Resource Allocation,PL2}[.45\linewidth][c]{%
    \includegraphics[width=\linewidth]{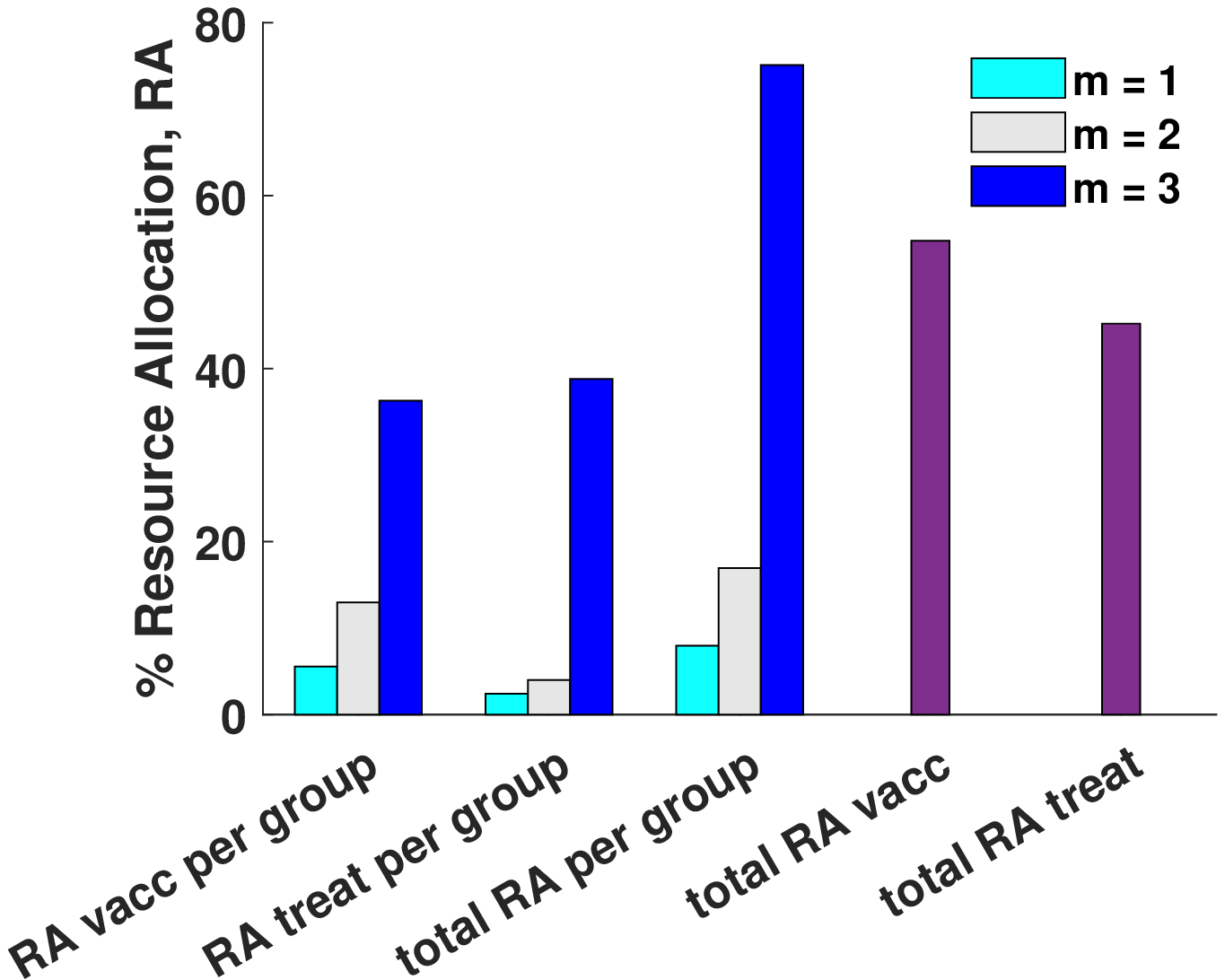}}
  \caption{{\small Controls and $\% $ Resource Allocation for ER and PL2 networks. Parameters: Spreading rate $ \beta = 0.5$, recovery rate $\gamma = 0.25$, epidemic duration $T = 20$, initial fraction of seed nodes $i_{0} = 0.01$, cost of vaccination $b = 0.25$, cost of treatment $c = 0.5. $}}
  \label{fig:Fig.4}
\end{figure*}
Shapes of the vaccination and treatment control signals for both networks are shown in Figs. 4a and 4b. Further, percentage allocation of resource in each of the $M=3$ groups for both the vaccination and treatment strategies are shown in Figs. 4c and 4d. Note that for Figs. 4c and 4d, the total resource adds up to $100\%$, that is, the first six bars that show resource allocation over vaccination in the three groups and resource allocation over treatment in the three groups add up to $100\%$. Similarly, bars seven to nine, which show total (vaccination plus treatment) resource allocated to each of the three groups, add up to $100\%$. And, bars ten and eleven, which show total (all three groups combined) resource allocated over the two control strategies, add up to $100\%$.

As seen from the plots in Figs. 4a and 4b, for the optimal strategy, the control signals assume larger strength in the early stages of the epidemic compared to the later stages in both the networks. In the early stages, the epidemic is known to grow exponentially making it the most critical stage in the spreading process. More number of infected individuals at the early stage is going to make the epidemic outbreak worse. The optimal strategy tries to either preempt infection by employing vaccines or remove these infected individual as early as possible. Hence, the strengths of vaccination and treatment control signals are greater in the early stages of the epidemic. Through fast early vaccination, the susceptible individuals gain immunity and are transferred to recovered state (without getting infected) thereby preventing the epidemic to progress further. Also, strong treatment control accelerates recovery, thereby minimizing the impact of infected individuals in the network. Therefore, it is worthwhile to apply the vaccination and treatment signals with larger strength at the early stages of the epidemic.

Figs. 4c and 4d identifies the relative importance of the three groups ($\mathcal{G}_{1}, \mathcal{G}_{2}, \mathcal{G}_{3}$) and the two control strategies for the two networks. These figures show that percentage of the total resource (vaccination plus treatment) taken up by low ($\mathcal{G}_{1}$), medium ($\mathcal{G}_{2}$) and high ($\mathcal{G}_{3}$) groups in the case of ER network is: $ 26.2 \%, 39.66 \% $, and $ 34.14 \% $; for PL2 network: $ 7.96 \%, 16.94 \% $, and $ 75.1 \% $ respectively. Thus, we can conclude that the ER network has a more homogeneous allocation of resource across the groups compared to the PL2 network. This is consistent with the nature of the degree distribution of the two networks. The allocation also shows that for the ER network, medium group $\mathcal{G}_{2}$ attracts more resources followed by high group $\mathcal{G}_{3}$ and low group $\mathcal{G}_{1}$. But, when the network topology is changed to PL2, maximum resource is allocated to the high group $\mathcal{G}_{3}$ followed by medium group $\mathcal{G}_{2}$ and low group $\mathcal{G}_{1}$. Not only this, high group $\mathcal{G}_{3}$ attracts disproportionately high allocation of resources in the case of the PL2 network. Due to disproportionate spreading potential of this group in the PL2 network, the infection would spread rapidly if these nodes get infected. To avoid this situation whatever the cost of vaccination and treatment, the optimal strategy tries to pre-empt infection by employing vaccines or remove these high degree infected nodes so as to prevent further infection.

Percentage of the total resource allocated to vaccination and treatment strategy in the case of the ER network are: $ 78.95 \% $ and $ 21.05 \% $ respectively; and for the PL2 network are: $ 54.8 \% $ and $ 45.2 \% $ respectively. This shows that the vaccination strategy is more important for the homogeneous network ER whereas treatment also assumes importance in the heterogeneous network PL2. That is, for the homogeneous network---which is devoid of high degree hubs or super-spreaders---the pre-emptive vaccination strategy is emphasized in the optimal strategy, where as, in presence of the high degree hubs (super-spreaders) in the heterogeneous networks, the treatment of such individuals also assumes importance. This is evident from the allocation of treatment resources to the high group ($\mathcal{G}_{3}$) for the PL2 network in Fig. 4d. Due to the large number of contacts, many of the high degree nodes get infected---vaccinations can only be carried out at a finite rate but a sizable population of the high degree hubs are present in the scale free network. Thus, treatment too assumes importance in the scale free networks.

\begin{figure*}[t]
  \centering
  \subcaptionbox{\small Cheap controls, $ b = 0.1, c = 0.1 $.}[0.45\linewidth][c]{%
    \includegraphics[width=\linewidth]{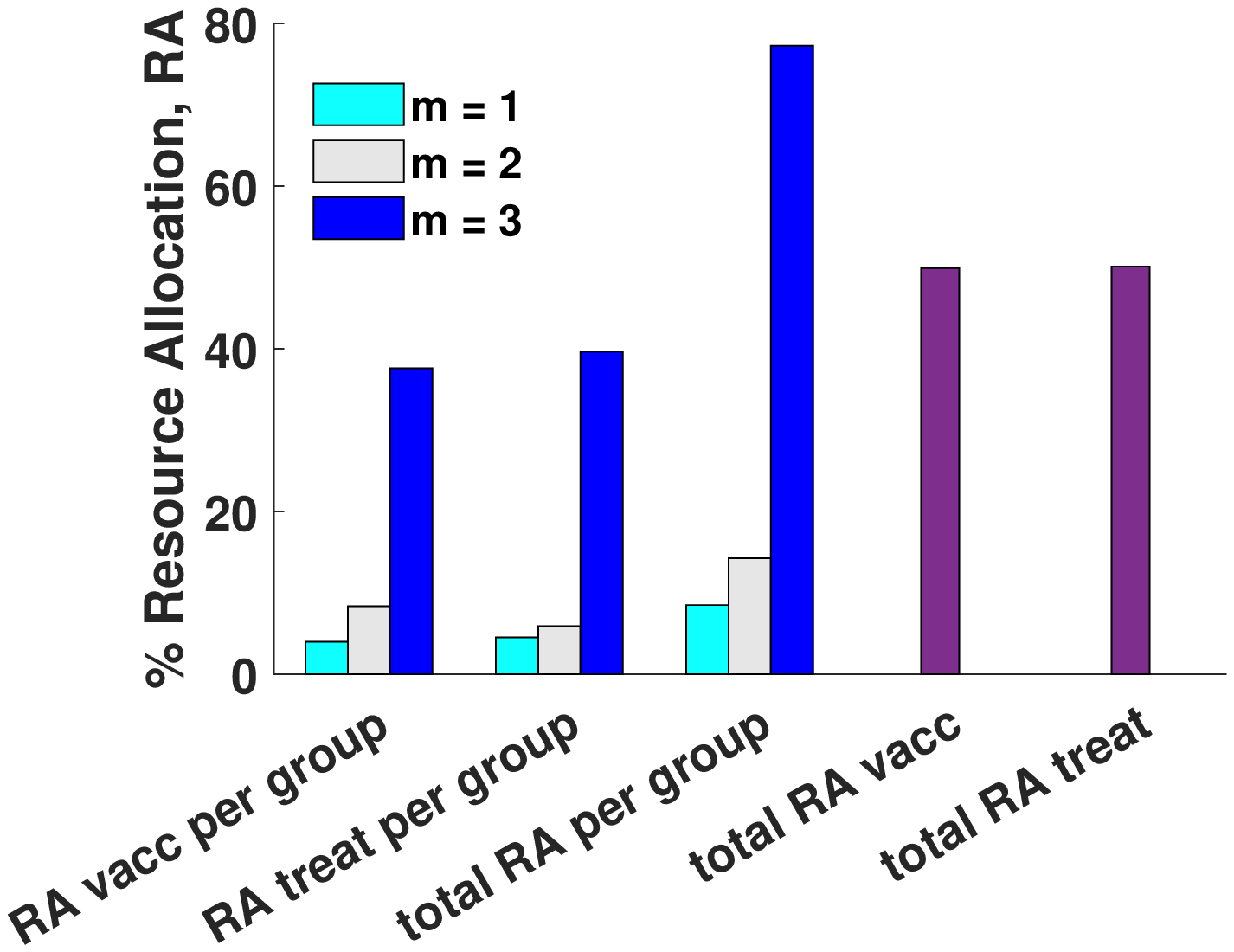}}
  \subcaptionbox{\small Expensive controls, $ b = 1, c = 1 $.}[0.45\linewidth][c]{%
    \includegraphics[width=\linewidth]{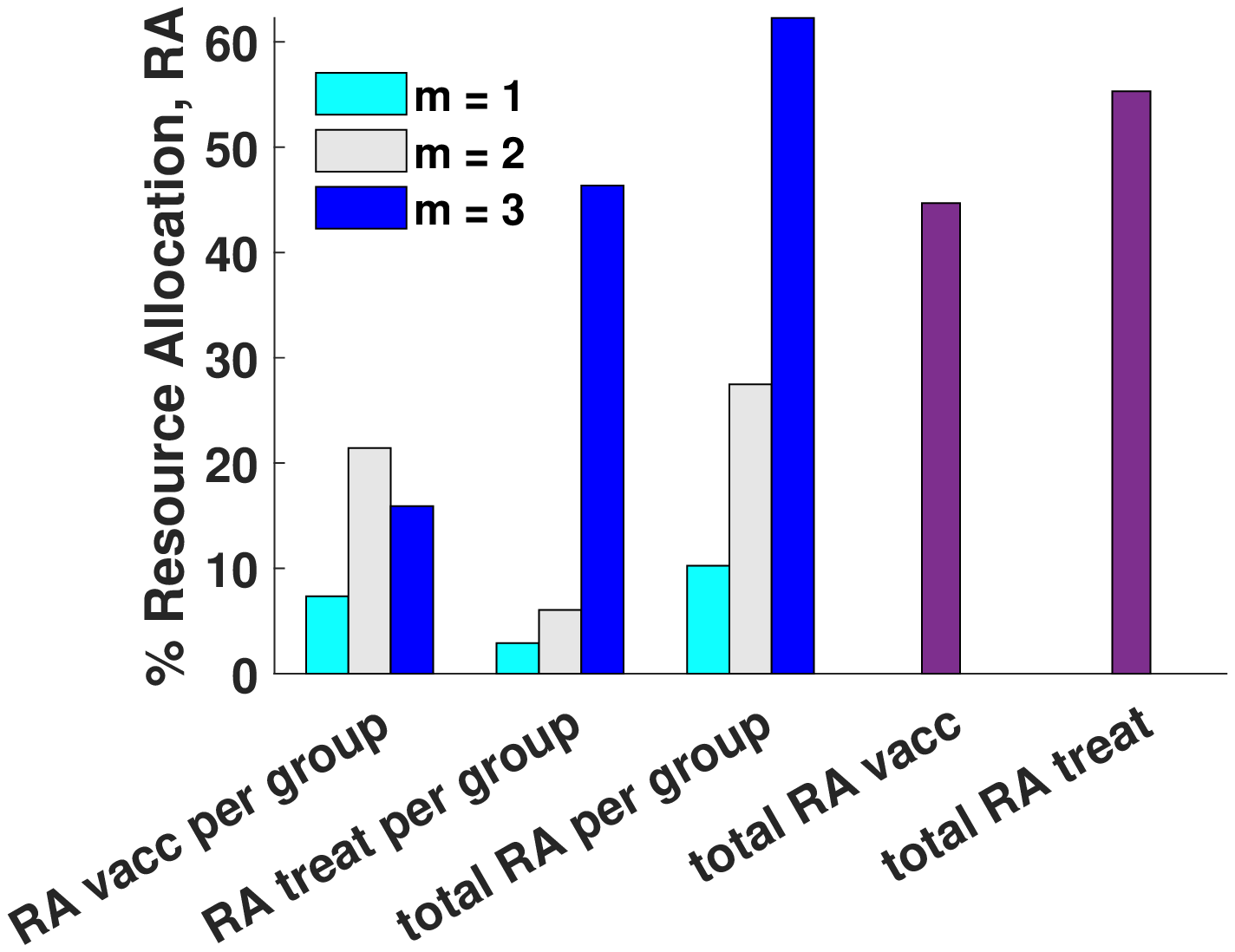}}
    \bigskip
    
  \subcaptionbox{\small Vaccination expensive than treatment, $ b = 0.5, c = 0.1 $.}[.45\linewidth][c]{%
    \includegraphics[width=\linewidth]{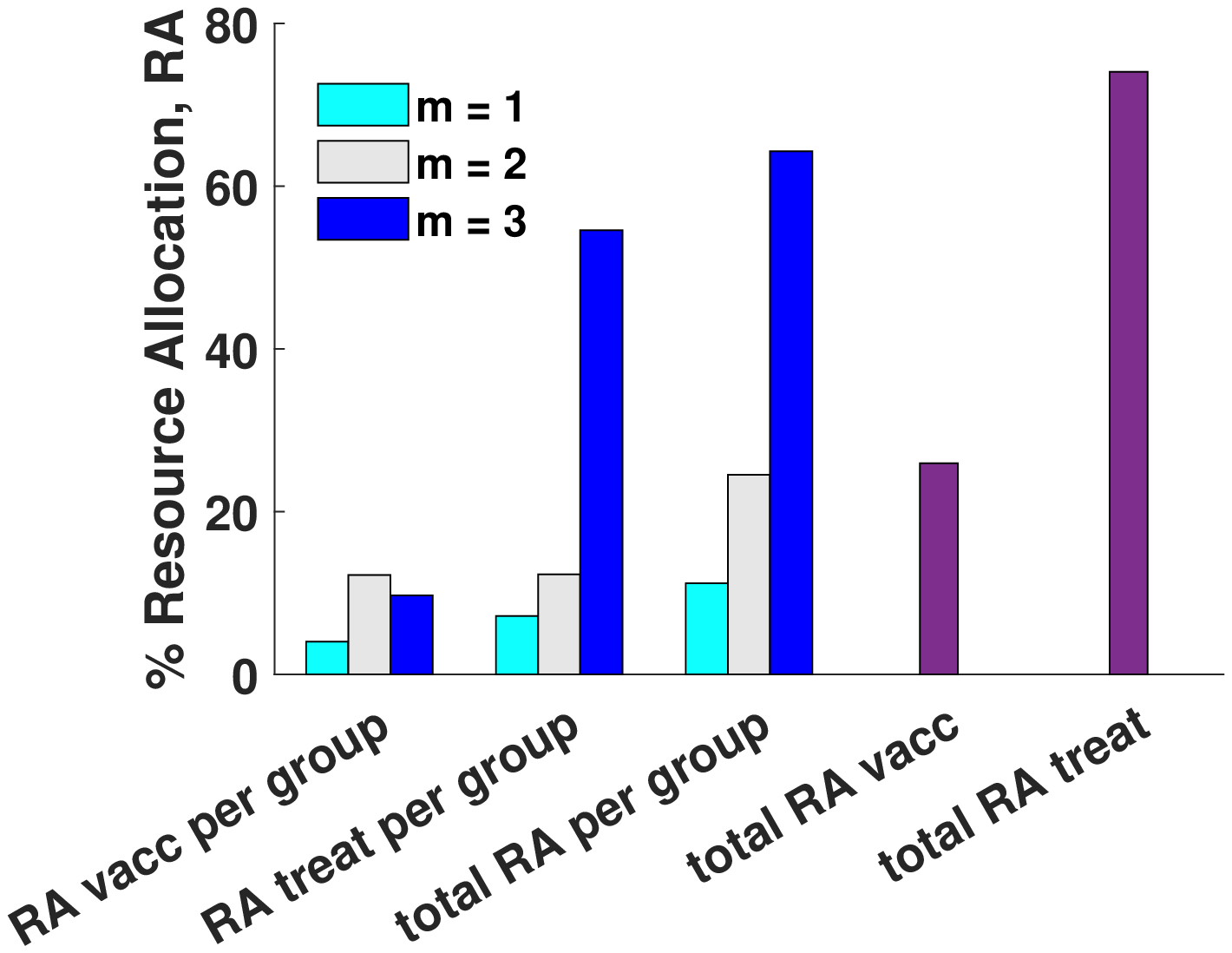}}
  \subcaptionbox{\small Treatment expensive than vaccination, $ b = 0.1, c = 0.5 $.}[.45\linewidth][c]{%
    \includegraphics[width=\linewidth]{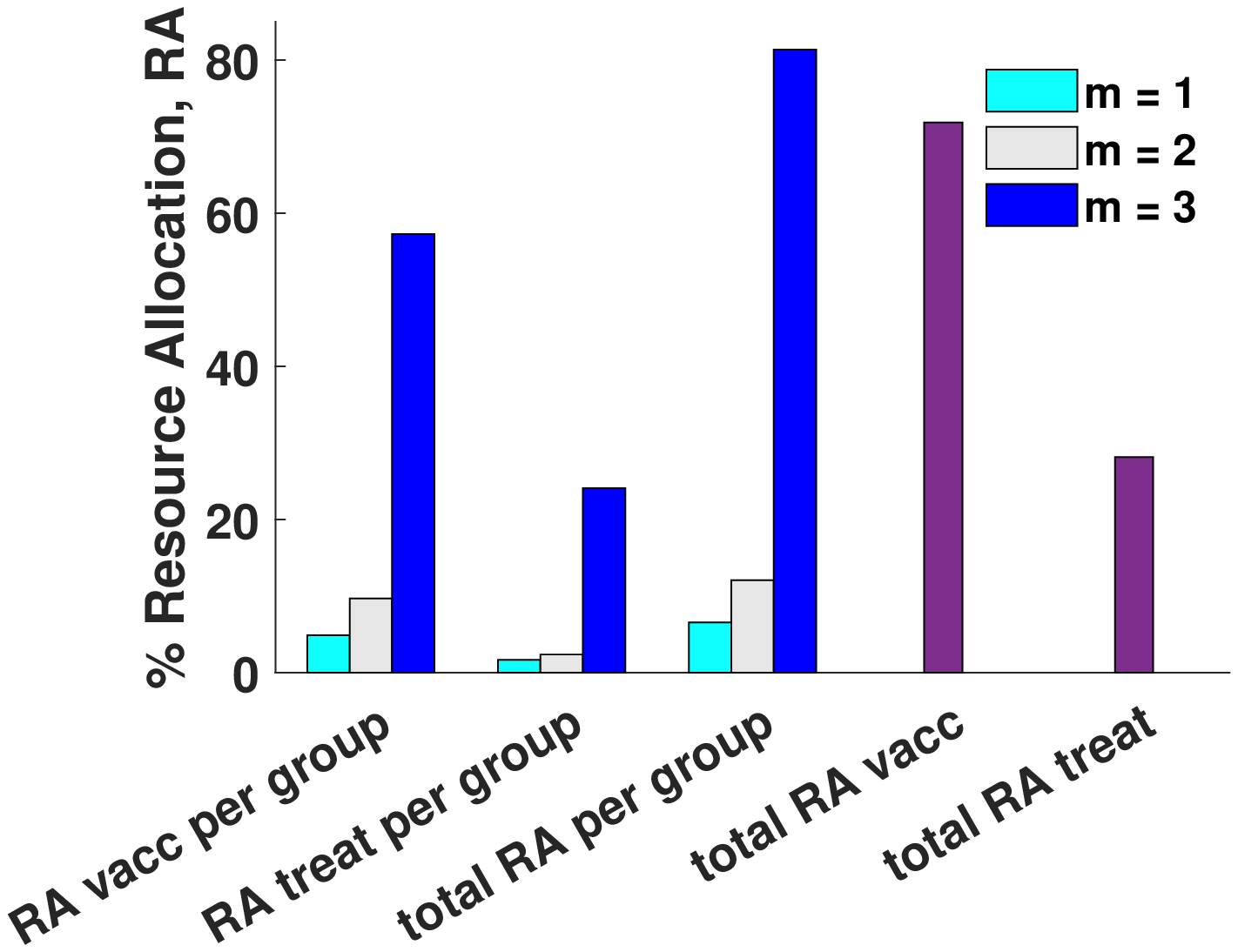}}
  \caption{{\small $\% $ Resource Allocation PL2 networks. Parameters: Spreading rate $ \beta = 0.5$, recovery rate $\gamma = 0.25$, epidemic duration $T = 20$, fraction of initial seed nodes $i_{0} = 0.01$. Cost of vaccination $b$, and cost of treatment $c$, vary as shown above.}}
  \label{fig:Fig.5}
\end{figure*}

In Fig. 5, we will see how the importance of groups $ \mathcal{G}_{m}, 1 \leq m \leq M $ and strategies change as the cost of control signals change. We have presented the results for the PL2 network only. If both vaccination and treatment control signals are cheap (Fig.5a) then, we see that both strategies are equally important in controlling the epidemic; with $ 49.9\% $ and $ 50.1\% $ of the total resource allocated to vaccination and treatment respectively. If both controls are expensive (Fig.5b) then the optimal strategy tries to allocate more resources in treating nodes from the high group $ (\mathcal{G}_{3}) $, followed by vaccinating nodes from the medium group $ (\mathcal{G}_{2}) $; with total resource allocated to vaccination and treatment being $ 44.7\% $ and $ 55.3\% $ respectively. If vaccination is more expensive than treatment (Fig.5c) then the optimal strategy allocates more resources ($ 74.06\% $) to treatment than vaccination ($ 25.94\% $), which is intuitive. Since vaccination strategy is expensive, individuals in $ (\mathcal{G}_{3}) $ who get infected are directly treated because treatment is cheaper. To mitigate the epidemic, the treatment strategy is exploited. Whereas, if treatment is more expensive than vaccination (Fig.5d) then, optimal strategy allocates $ 71.85\% $ of resource to vaccination and $ 28.15\% $ of resource to treatment. The trend of $ High > Medium > Low $ in terms of percentage resource allocation among groups is observed for both the strategies. 

\subsection{Effect of the Spreading Rate, $ \beta $}

\begin{figure*}[h]
  \centering
  \subcaptionbox{\small Objective function $ J $ vs. spreading rate $ \beta $, ER}[.45\linewidth][c]{%
    \includegraphics[width=\linewidth]{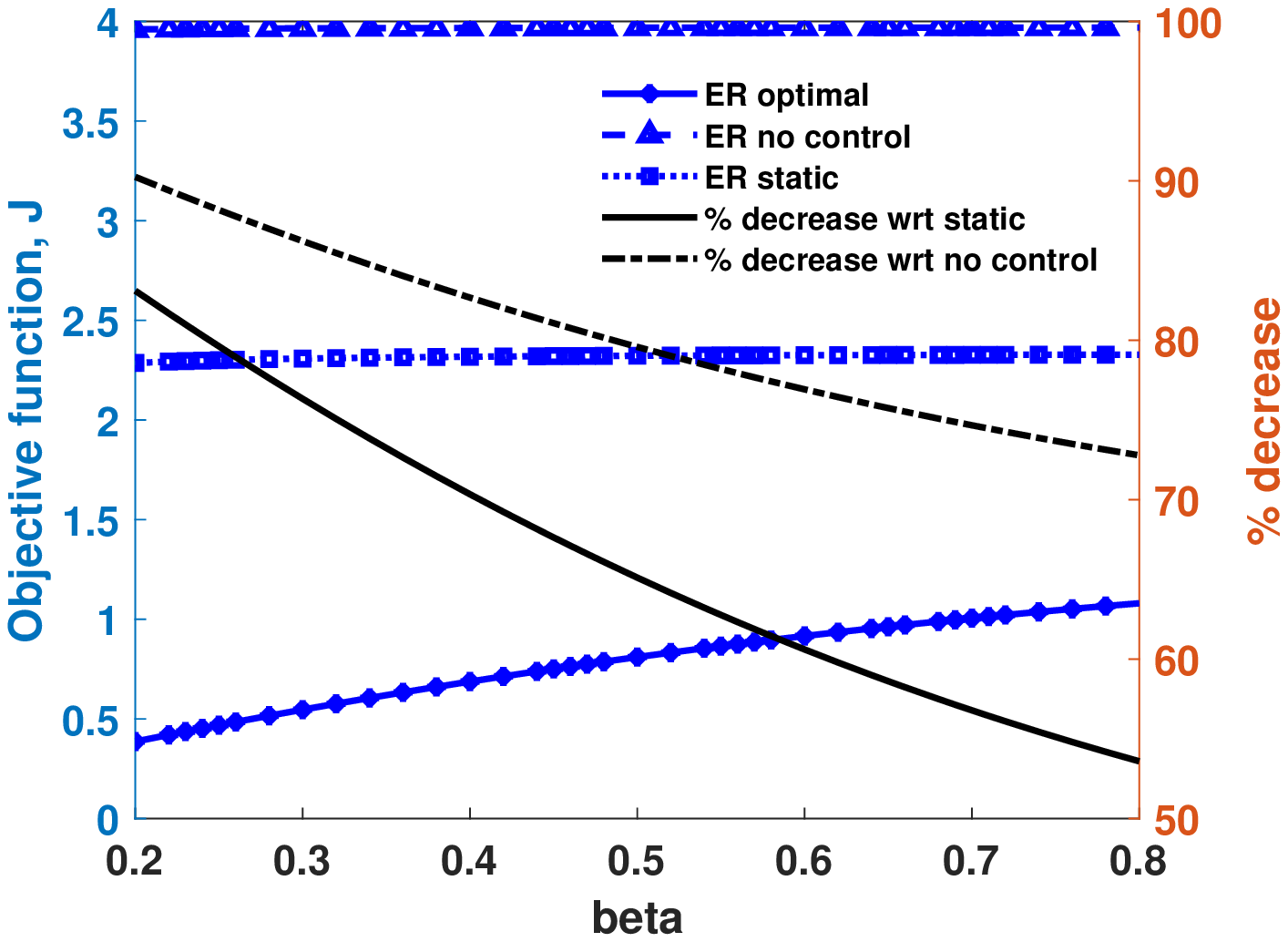}}
  \subcaptionbox{\small Objective function $ J $ vs. spreading rate $ \beta $, PL2}[.45\linewidth][c]{%
    \includegraphics[width=\linewidth]{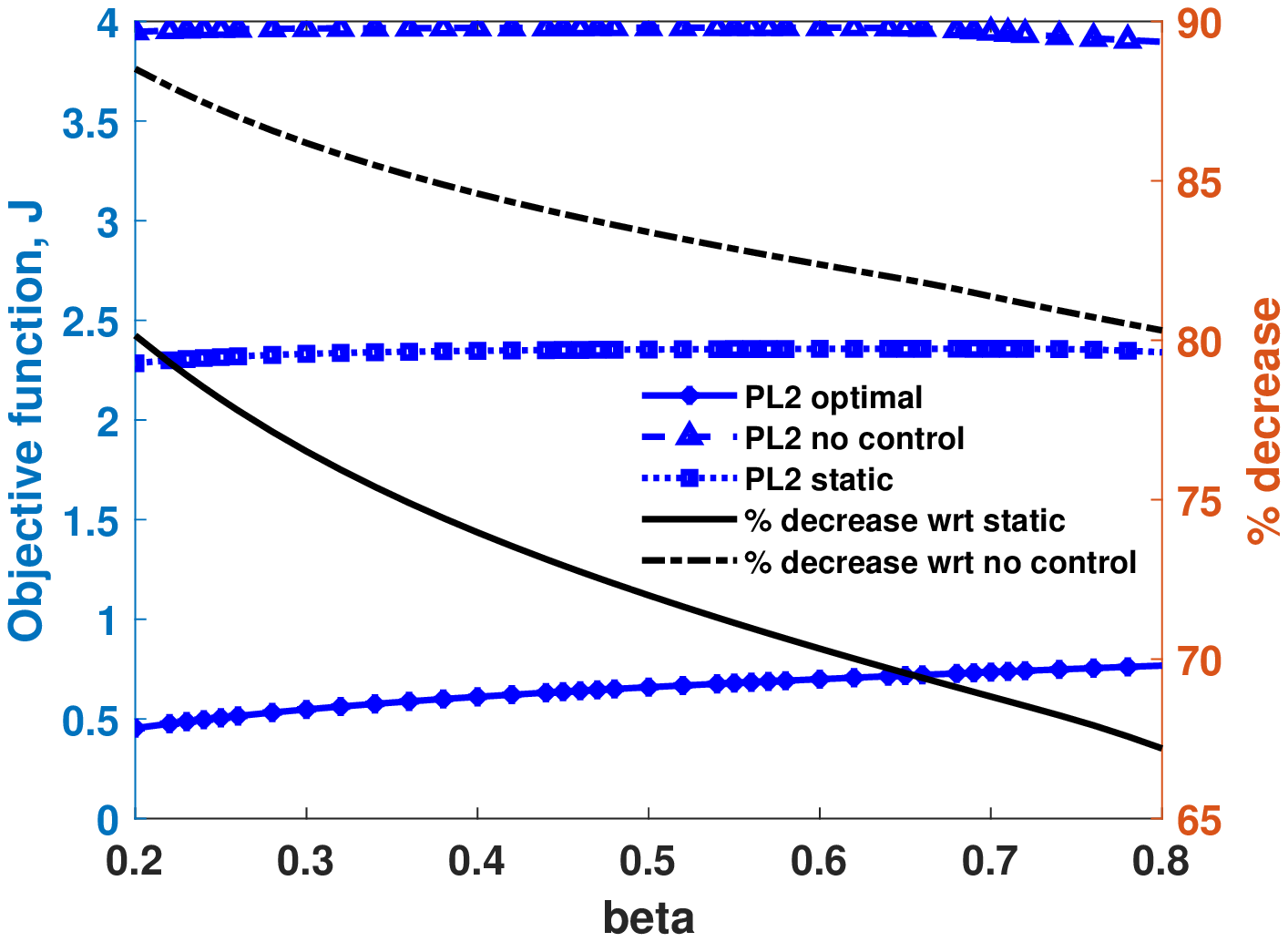}}
  \bigskip

  \subcaptionbox{\small Cumulative infected population vs. spreading rate $ \beta $ , ER}[.45\linewidth][c]{%
    \includegraphics[width=\linewidth]{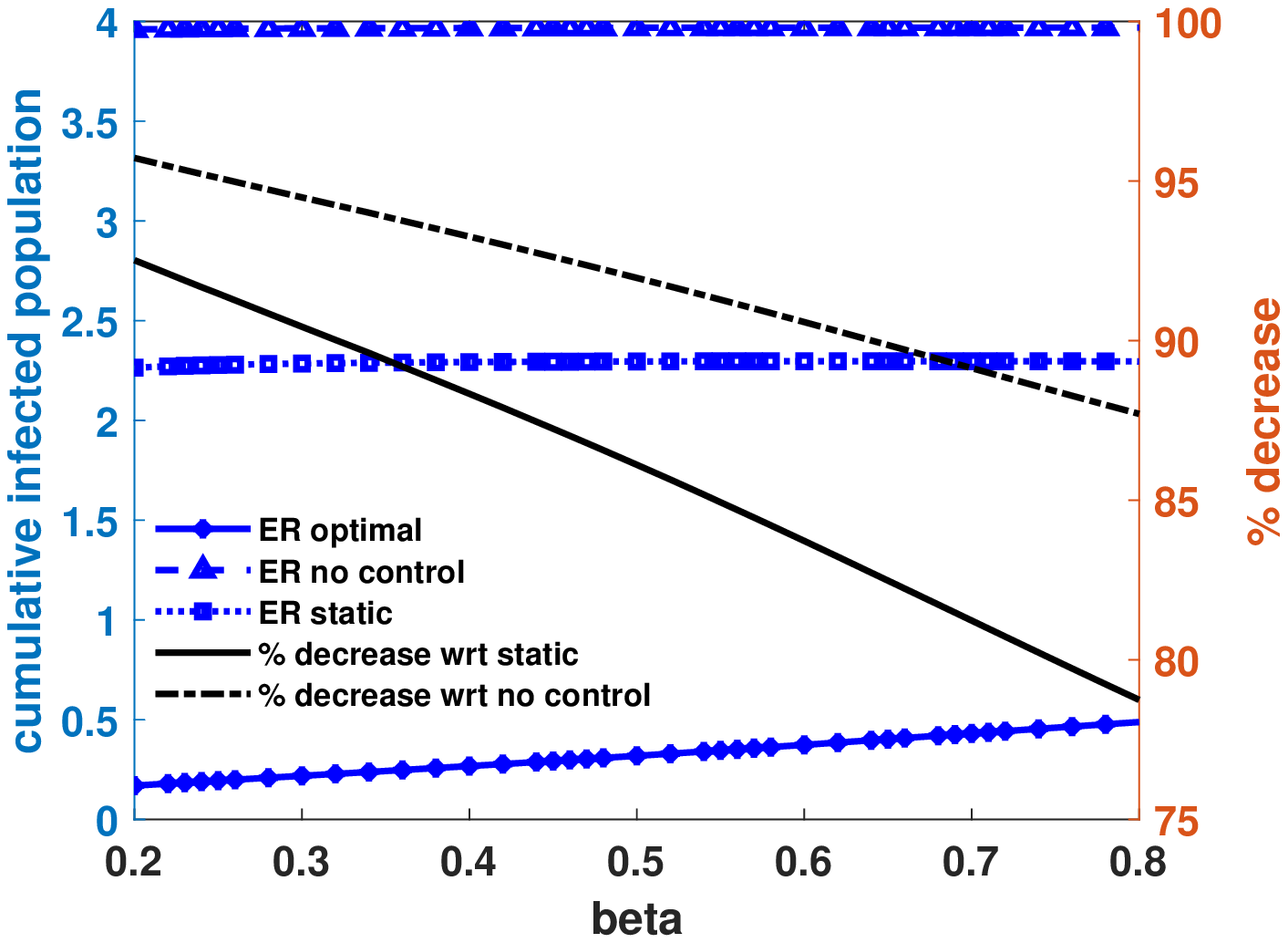}}
  \subcaptionbox{\small Cumulative infected population vs. spreading rate $ \beta $ , PL2}[.45\linewidth][c]{%
    \includegraphics[width=\linewidth]{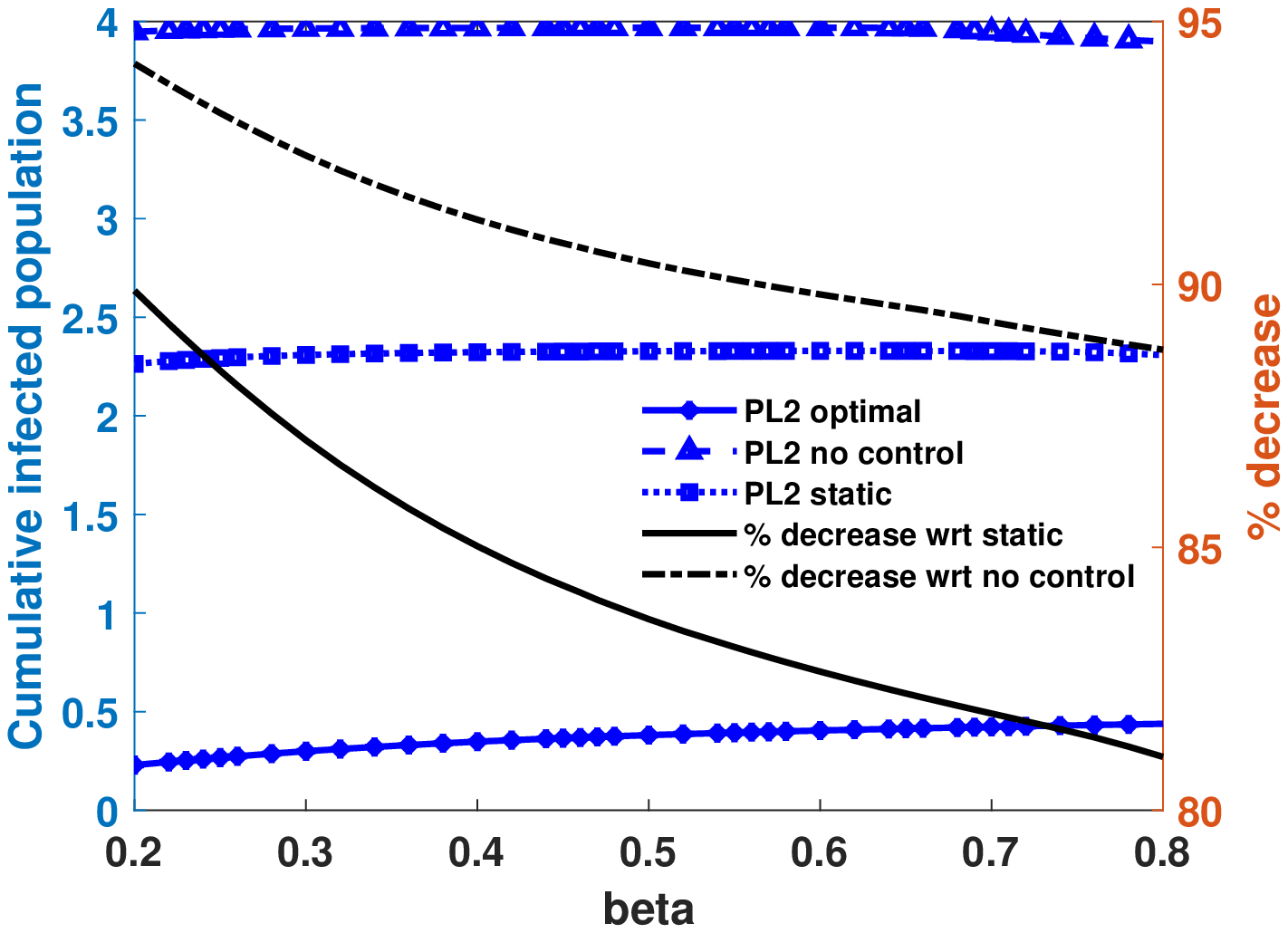}}
  \caption{{\small Variation of objective function $ J $ and cumulative infected population with respect to spreading rate $\beta $ for ER and PL2 networks. Parameters: Recovery rate $ \gamma = 0.25$, epidemic duration $T = 20$, initial fraction of seed nodes $i_{0} = 0.01$, cost of vaccination $b = 0.25$, cost of treatment $c = 0.5. $}}
  \label{fig:Fig.6}
\end{figure*}

The effect of varying the spreading rate, $ \beta $ on the objective function, $ J $ is studied in Figs. 6a, and 6b. The optimal control strategy minimizes the objective (cost) function, $ J $ given by Eq.(4). We observe that the optimal strategy is able to achieve more than $ 65\% $ improvement in the value of $ J $ for the PL2 network, and more than $ 50\% $ improvement for the ER network, over the constant (static) control heuristic strategy. As expected, the optimal strategy always works better than the non-optimal strategies. From Figs. 6a, 6b we can see that the values of $ J $ for the non-optimal strategies are almost the same as we vary $ \beta $ from $ 0.2 $ to $ 0.8 $. However, there is a significant reduction in the values of $ J $ for all $ \beta $ for the optimal strategy. Even for high spreading rate ($ \beta = 0.8 $), the optimal strategy is able to achieve a percentage improvement of $ 67.2\%$ and $80.31\% $ compared to the constant control and no control strategies for the PL2 network; $ 53.59\%$ and $72.78\% $ compared to the constant control and no control strategies for the ER network.

We also observe that a more virulent epidemic is less amenable to control, even with the optimal strategy, than a less virulent epidemic. This is consistent with real world experience. A less infectious disease is usually manageable quite easily compared to a highly infectious disease (like the current Covid-19 pandemic which has wreaked havoc throughout the world). As epidemic becomes more virulent (that is, the spreading rate $\beta$ for the epidemic increases), susceptible individuals become infected at an increasing rate during the phase when epidemic is on the rise. To counter this, we need controls of larger strengths to shift individuals to the recovered class---from the susceptible class through vaccination, and from the infected class through treatment controls. However, stronger controls are more costly to implement because of the convex increasing nature of the cost function considered in this paper, which leads to this trend. 

For a wide range of the spreading rate $\beta$, the optimal strategy achieves a significant percentage improvement in the objective function $J$ when compared to the non optimal strategy. Hence, it is important to calculate and implement optimal strategies irrespective of whether the epidemic is more virulent or less (however, the advantage provided by the optimal management is more pronounced for less virulent epidemic as discussed above). The objective function $ J $ (Eq. 4) includes both monetary benefits and the health state of the population. When controlling an epidemic, monetary benefits in terms of resource allocation are important for the government and the healthcare authorities. While the general population is more concerned with their health. Hence, we present the cumulative infected population (which is the first part of Eq. (4) and represent the health state of the population) as we vary $ \beta $, in Figs. 6c and 6d. From the figures we can see that there is significant reduction in the cumulative infected population in the optimal strategy compared to the non-optimal heuristics. 
  
\subsection{Effect of Changing The Relative Cost of Control Signals}

\begin{figure*}[t!]
  \centering
  \subcaptionbox{\small ER}[0.45\linewidth][c]{%
    \includegraphics[width=\linewidth]{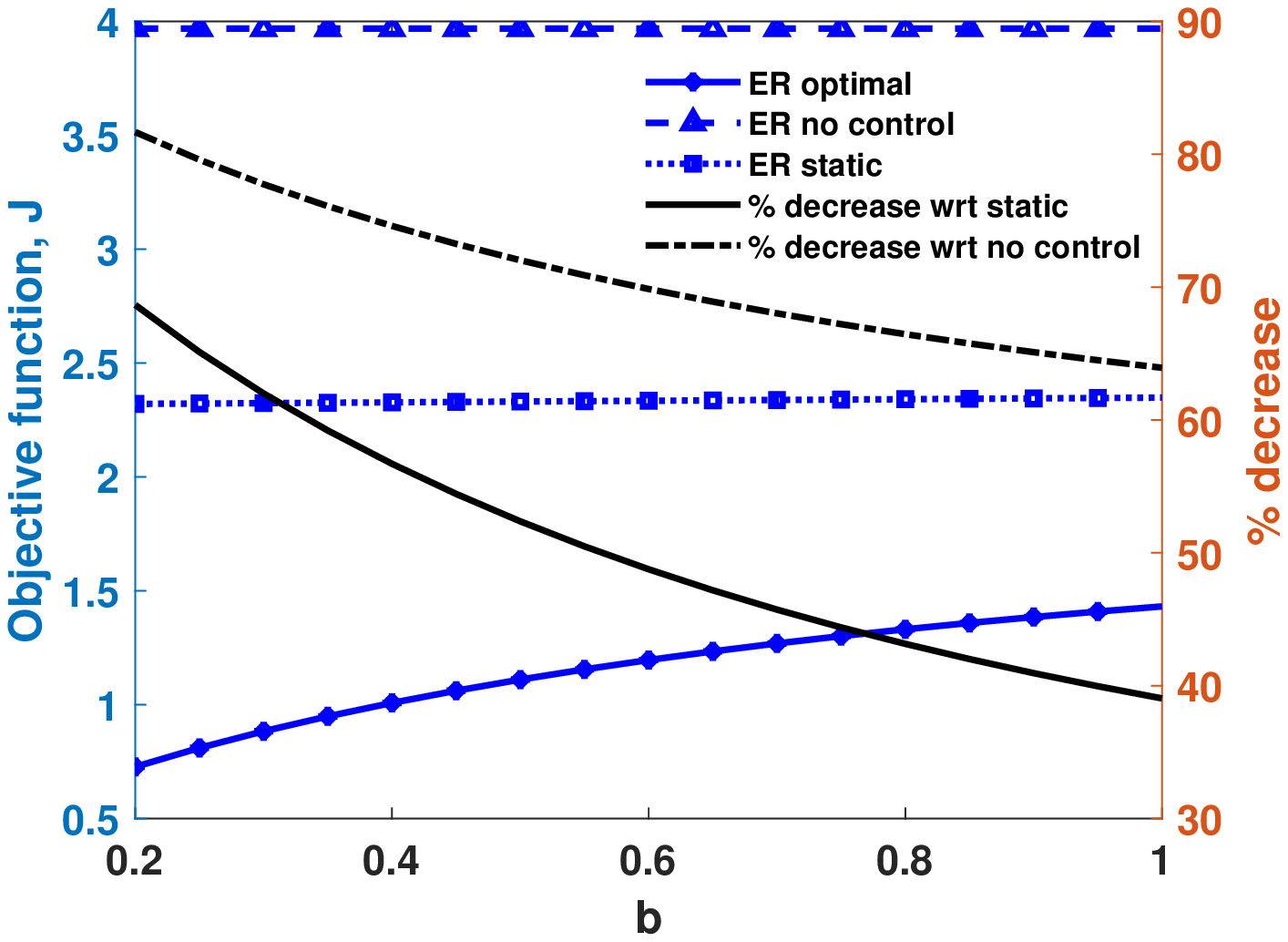}}\quad
  \subcaptionbox{\small PL2}[.45\linewidth][c]{%
    \includegraphics[width=\linewidth]{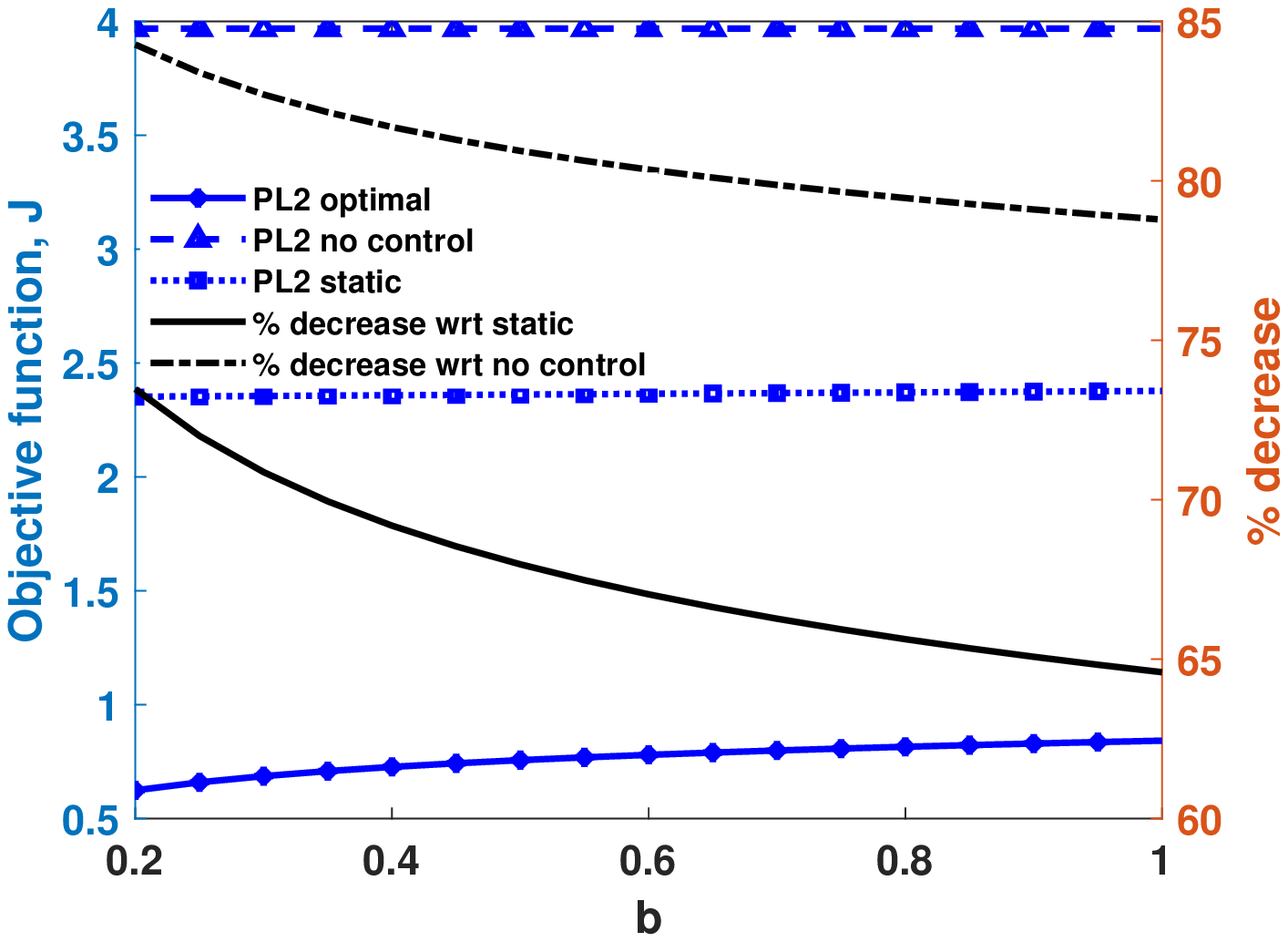}}\quad
  \caption{{\small Objective function $ J $ vs. cost of vaccination $b$. Parameters: Spreading rate $ \beta = 0.5$, recovery rate $\gamma = 0.25$, epidemic duration $T = 20$, initial fraction of seed nodes $i_{0} = 0.01$, cost of treatment $c = 0.5. $}}
  \label{fig:Fig.7}
\end{figure*}

\textit{Vaccination Control Signal:} The effect of varying the cost of vaccination control signal is captured by parameter $ b $ in the cost function (4) and is studied in Fig. 7. For the PL2 network, the value of $ J $ at lower vaccination cost, $ b = 0.2 $ is $ 0.6241 $ and at higher vaccination cost, $ b = 1 $ is $ 0.8419 $. Similarly, for the ER network, the value of $ J $ at lower vaccination cost, $ b = 0.2 $ is $ 0.7273 $ and at higher vaccination cost, $ b = 1 $ is $ 1.4314 $. From above data we can see that there is approximately $ 2 $ fold increase in the value of $ J $ for ER network and approximately $ 1.35 $ fold increase for PL2 network, as we vary $ b $ from $ 0.2 $ to $ 1 $. Increasing the cost of vaccination hurts the ER network more than the PL2 network because vaccination is more important in mitigating the epidemic for the ER network. Also, as one would expect, as the cost of vaccination increases, the percentage improvement achieved by the optimal strategy over the heuristics decreases for both the networks.

\begin{figure*}[h]
  \centering
  \subcaptionbox{\small ER}[.45\linewidth][c]{%
    \includegraphics[width=\linewidth]{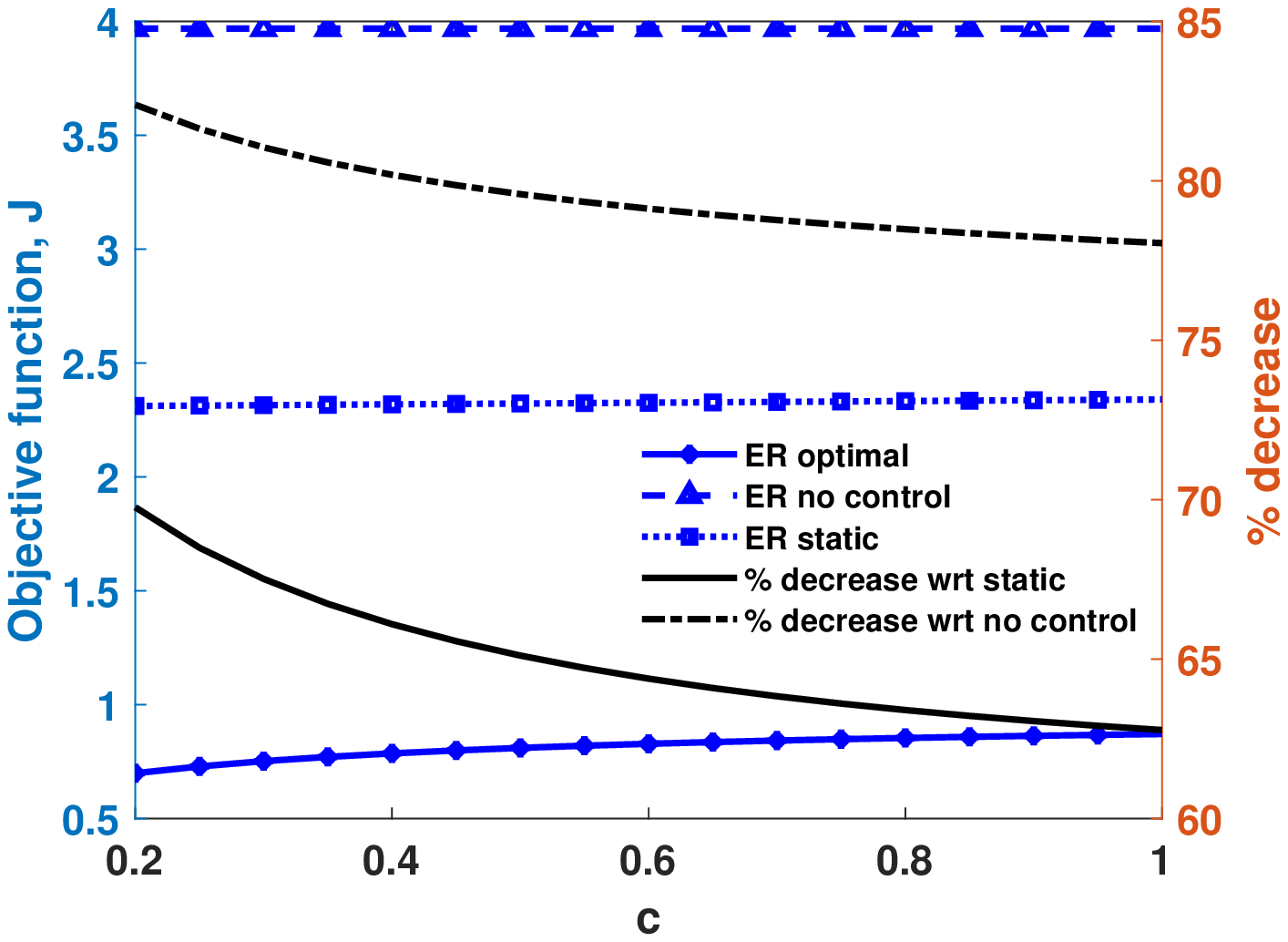}}\quad
  \subcaptionbox{\small PL2}[.45\linewidth][c]{%
    \includegraphics[width=\linewidth]{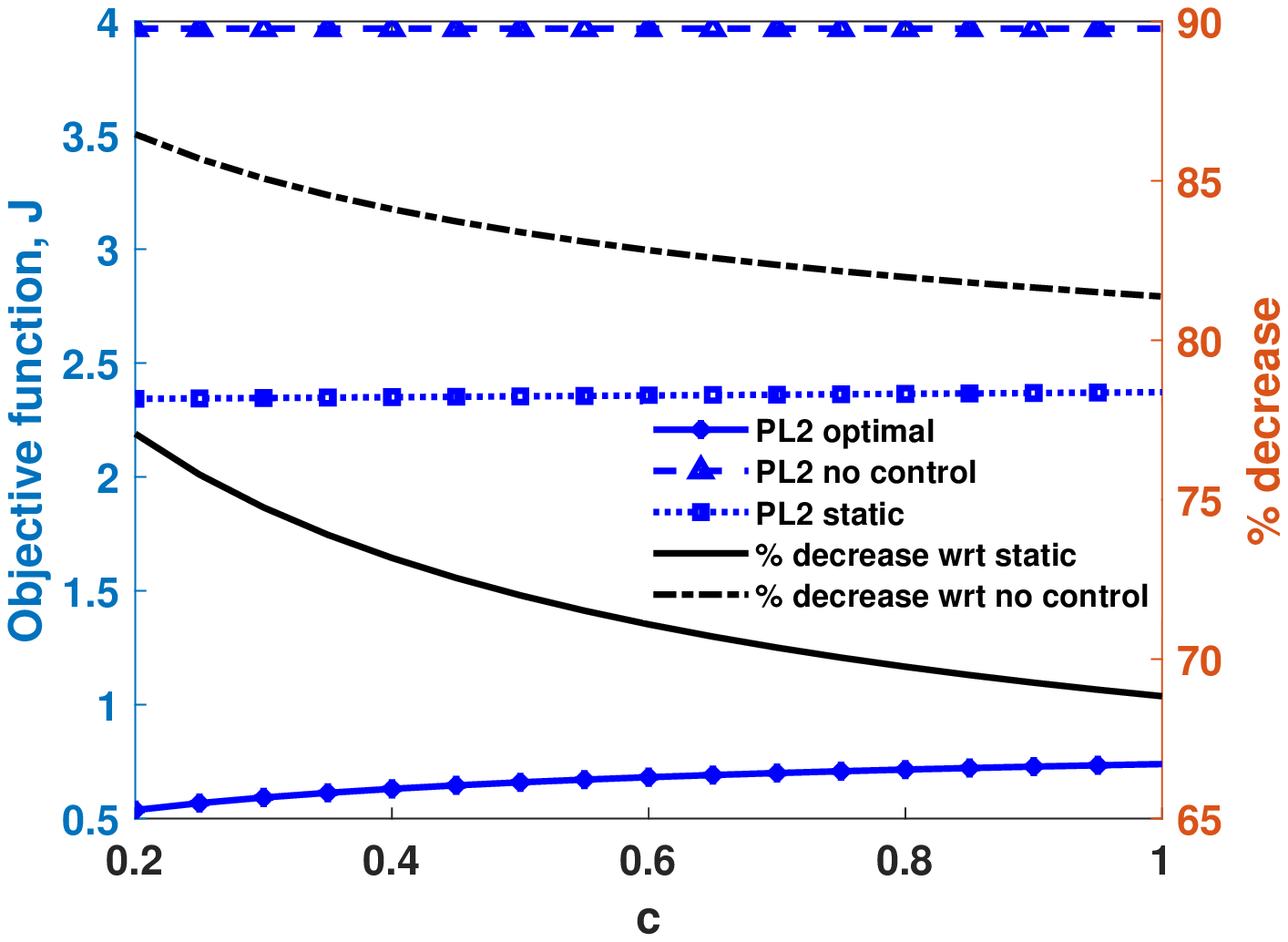}}\quad
  \caption{{\small Objective function $ J $ vs. cost of treatment $ c $. Parameters: Spreading rate $ \beta = 0.5$, recovery rate $\gamma = 0.25$, epidemic duration $T = 20$, initial fraction of seed nodes $i_{0} = 0.01$, cost of vaccination $b = 0.25. $}}
  \label{fig:Fig.8}
\end{figure*}

\textit{Treatment Control Signal:} Fig. 8 shows the effect of varying the cost of treatment control signal, which is captured by parameter $ c $ in the cost function (4). For the PL2 network, the value of $ J $ at lower treatment cost, $ c = 0.2 $ is $ 0.5873 $ and at higher treatment cost, $ c = 1 $ is $ 0.7392 $. Similarly, for the ER network, the value of $ J $ at lower treatment cost, $ c = 0.2 $ is $ 0.6989 $ and at higher treatment cost, $ c = 1 $ is $ 0.8711 $. From above data we can see that there is approximately $ 1.25 $ fold increase in the value of $ J $ for the ER network and approximately $ 1.26 $ fold increase for the PL2 network, as we vary $ c $ from $ 0.2 $ to $ 1 $. Thus, increasing treatment cost has less impact on performance of the optimal strategy for both the networks compared to the vaccination cost.

In this section we notice that in-spite of a five fold increase in the cost of vaccination or treatment, the optimal strategy achieves a significant percentage improvement in the value of cost function over the non optimal strategies. This is similar to what we saw in Section 4.2 for the case of spreading rate, and demonstrates that optimal strategies are important to implement over a wide range of system parameters.

\subsection{Real World Network}
\begin{figure*}[t!]
	\centering
	\subcaptionbox{\small Controls for real world network}[0.45\linewidth][c]{%
		\includegraphics[width=\linewidth]{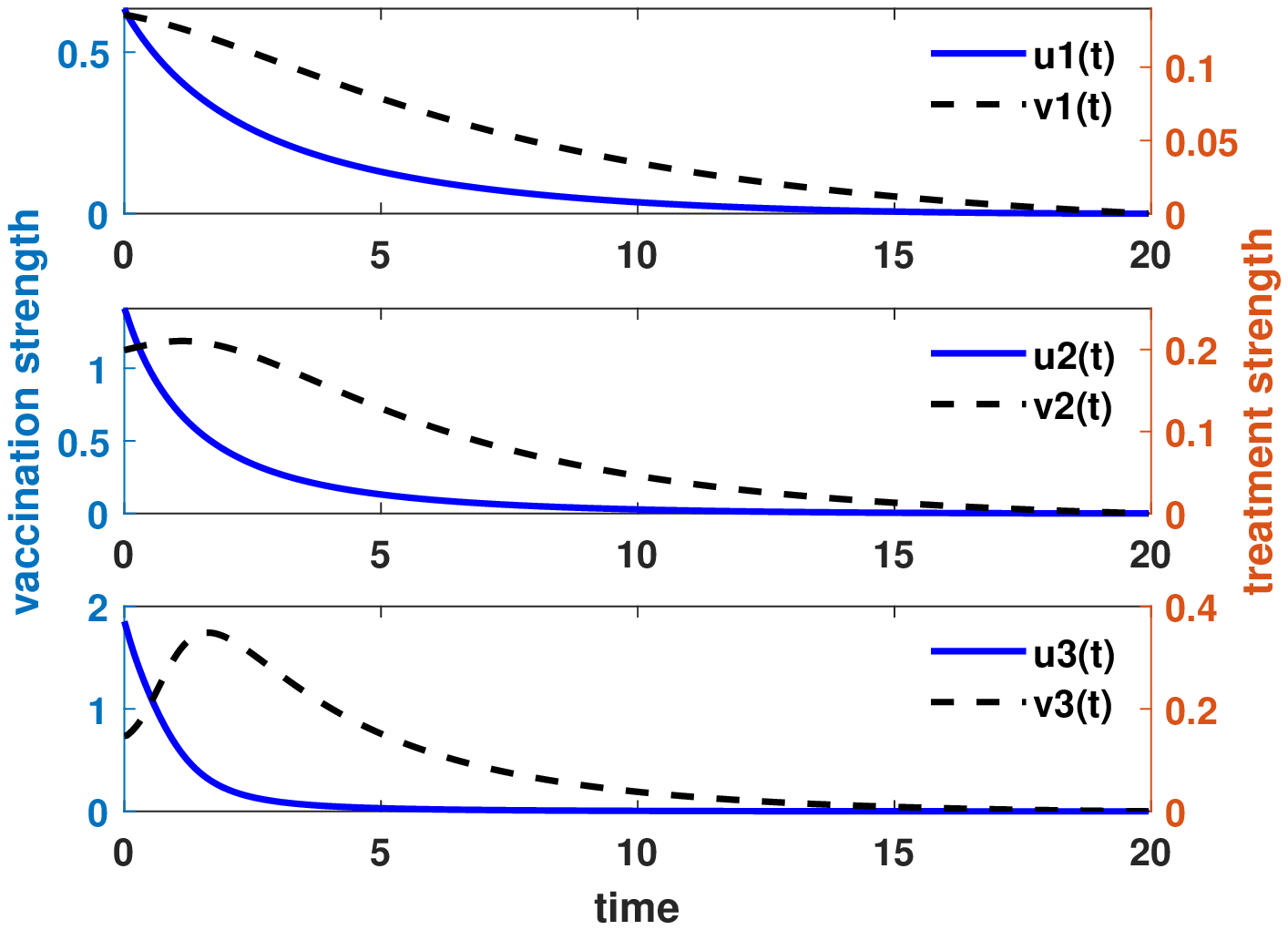}}\quad
	\subcaptionbox{\small \% resource allocation for real world network}[.45\linewidth][c]{%
		\includegraphics[width=\linewidth]{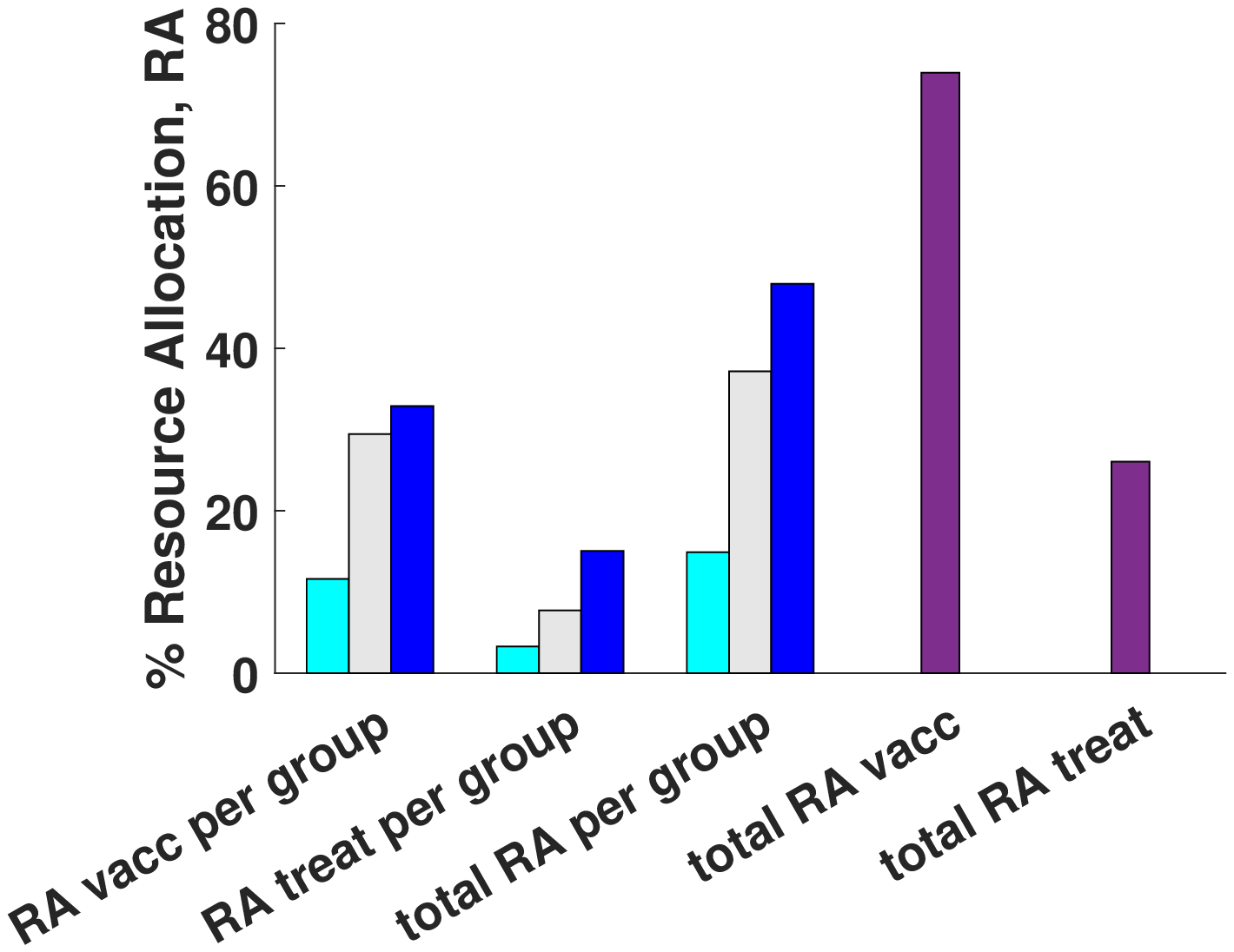}}\quad
	\caption{{\small Controls and \% resource allocation for real world network. Parameters: Spreading rate $ \beta = 0.5$, recovery rate $\gamma = 0.25$, epidemic duration $T = 20$, initial fraction of seed nodes $i_{0} = 0.01$, cost of vaccination $b = 0.25$, cost of treatment $c = 0.5. $}}
	\label{fig:real_world_nw}
\end{figure*}

Fig. \ref{fig:real_world_nw} presents results for a publicly available real world network that was specifically obtained to study transmission of infectious diseases \cite{real_world_nw}. The network captures close proximity interactions of 788 members of a high school (655 students, 73 teachers, 55 staff members, and 5 others) over the period of a day. Only interactions that are six minutes or greater are considered. The network has $|\mathcal{K}| = 47$ degree classes that are divided into $Z=21$ groups, and further into $M=3$ groups of groups, using the method provided in Sections 2.2 and 2.3 respectively. The mean degree of the network is 17.86.

The nature of controls in this real world network (Fig. \ref{fig:real_world_nw}a) are similar to what we saw in the ER and PL2 networks (Figs. 3a, 3b)---the control effort gets concentrated in the beginning of the epidemic period. The resource allocation over the $M=3$ groups (Fig. \ref{fig:real_world_nw}b) are similar to the resource allocation for the PL2 network (Fig. 3d) with the high degree group $\mathcal{G}_{3}$ attracting most resources, followed by the medium degree group $\mathcal{G}_{2}$, and finally the low degree group $\mathcal{G}_{1}$. However, there is one difference in the resource allocation plot here compared to that of the PL2 network: The vaccination strategy is allocated significantly more resources than the treatment strategy in the real world network (Fig. \ref{fig:real_world_nw}b), unlike in the PL2 network (Fig. 3d). This shows that the optimal strategy is to be proactive and immunize individuals through vaccination. 

The values of the cost function $J$ (Eq. 4) for the optimal, constant, and no control strategies are 0.7977, 2.3105, and 3.9569 respectively. The optimal strategy achieves a significant percentage improvement of 65.48\% and 79.84\% over the constant and no control strategies. Thus, the huge improvements obtained by the optimal setup over the heuristics in the synthetic networks (Figs. 6, 7, 8) are carried forward in the real world networks too.

\section{Conclusion}

We studied the problem of devising optimal strategies for vaccination and treatment for minimizing cumulative infected individuals (nodes), while at the same time minimizing the cost of control signals for mitigating the effects of a biological epidemic. We used the degree based \textit{Susceptible-Infected-Recovered} compartmental model. Nodes of the same degrees are assigned to the same degree class. We form $ Z $ groups of these degree classes, this reduces computational effort without affecting the performance much. We further divide these $ Z $ groups into $ M $ different groups (groups of groups) and apply separate vaccination and treatment control signals to each of these groups. We solved the optimal control problem using Matlab's nonlinear optimization solver \textit{fmincon()}. Our methodology works for a network with arbitrary degree distribution. We studied and presented the results for the \textit{Erd\H{o}s-R\'{e}nyi}, \textit{scale free} (power law exponent, $ \alpha = 2 $) and a real world network (available in \cite{real_world_nw}). We studied the effects of various model parameters on the optimal control strategy and quantified the improvement of the optimal strategy over the non-optimal control strategies: no control strategy and constant control strategy.
 
The solution of the optimal control problem provides the best resource allocation over the epidemic duration, the strategies, and the $ M $ groups, effectively mitigating the spread of the disease. We also identify important groups which should be allocated more resources for controlling the epidemic at the earliest. The order of importance of the groups is different for different network topologies. For the scale free network, the order importance is $ High > Medium > Low $. In contrast for the Erd\H{o}s-R\'{e}nyi network, it is $ Medium > High > Low $. We observe that for the Erd\H{o}s-R\'{e}nyi network, the vaccination control strategy plays a significant role in controlling the epidemic and hence is more important. Whereas, for the scale free network both strategies are approximately equally important. The real world network displays a behavior which is hybrid of the two---order of importance of the groups in resource allocation is $ High > Medium > Low $, but vaccination assumes more importance than treatment in the optimal strategy.

We also studied the effect of relative cost of vaccination and treatment on resources allocated to the different strategies and the $ M $ groups for the scale free network. If a particular strategy is more expensive than the other, then the cheaper strategy is exploited---more resources are allocated to the cheaper strategy. Our results show that the optimal strategy significantly outperforms the non optimal heuristic strategies, and therefore should be preferred over the heuristics, for all the networks studied. Our framework may be of interest to governments and healthcare authorities for devising effective vaccination and treatment strategies during an epidemic outbreak.

\section*{Acknowledgment}
This research was supported in part by SERB India via the grant SRG/2019/001054.

\end{document}